\documentclass[11pt,british]{article}
\usepackage[T1]{fontenc}
\usepackage[latin9]{inputenc}
\usepackage{geometry}
\usepackage{amsmath}
\usepackage{amssymb}
\usepackage{stackrel}
\usepackage{esint}

\makeatletter

\providecommand{\tabularnewline}{\\}

\usepackage{babel}

\makeatother

\usepackage{babel}
\begin{document}

\title{Roaming form factors for the tricritical to critical Ising flow }

\author{D. X. Horváth\thanks{esoxluciuslinne@gmail.com}$\;{}^{1,2}$, P.
E. Dorey\thanks{p.e.dorey@durham.ac.uk}$\;{}^{3}$ and G. Takács\thanks{takacsg@eik.bme.hu}$\;{}^{1,2}$\\
 ~\\
 $^{1}${\small{}MTA-BME \textquotedbl{}Momentum\textquotedbl{} Statistical
Field Theory Research Group}\\
 {\small{}1111 Budapest, Budafoki út 8, Hungary}\\
 ~\\
 $^{2}${\small{}Department of Theoretical Physics, }\\
{\small{} Budapest University of Technology and Economics}\\
{\small{}1111 Budapest, Budafoki út 8, Hungary}\\
 ~\\
 $^{3}${\small{}Department of Mathematical Sciences, }\\
{\small{}Durham University, }\\
{\small{}South Road, Durham DH1 3LE, United Kingdom}}

\date{1st July 2016}
\maketitle
\begin{abstract}
We study the massless flows described by the staircase model introduced
by Al.B. Zamolodchikov through the analytic continuation of the sinh-Gordon
S-matrix, focusing on the renormalisation group flow from the tricritical
to the critical Ising model. We show that the properly defined roaming
limits of certain sinh-Gordon form factors are identical to the form
factors of the order and disorder operators for the massless flow.
As a by-product, we also construct form factors for a semi-local field
in the sinh-Gordon model, which can be associated with the twist field
in the ultraviolet limiting free massless bosonic theory. 
\end{abstract}

\section{Introduction}

The staircase flow has attracted considerable interest since Al.B.
Zamolodchikov's intriguing preprint (later published as \cite{RefZamo})
appeared more than 20 years ago. In his work, Zamolodchikov observed
that the S-matrix of the sinh-Gordon quantum field theory (QFT) in
two dimensions can be analytically continued away from the self-dual
point of the model to complex values of the coupling in such a way
that the resulting S-matrix makes perfect sense as a scattering theory.
Due to the lack of a proper Lagrangian description, he used the thermodynamic
Bethe Ansatz (TBA) to study the properties of the analytically continued
model. The $c$-function obtained from the TBA equations showed a
peculiar behaviour as the real parameter $\theta_{0}$ encoding this
continuation increased: a `staircase' of more and more clearly defined
plateaux appeared, at heights equal to the central charges $c=1-6/p(p+1)$
of the conformal minimal models $\mathcal{M}_{p}$\,, and in the
intervals between the plateaux the flow was found to approximate the
crossovers $\mathcal{M}_{p}\rightarrow\mathcal{M}_{p-1}$ between
the minimal models generated by the integrable perturbing operator
$\phi_{1,3}$. This observation led to the interpretation of the model
in this so-called roaming limit as a field theory describing a renormalisation
group (RG) flow that passes by the successive conformal minimal models,
eventually ending in a trivial IR fixed point which corresponds to
a free massive Majorana fermion, i.e. the scaling Ising model. The
larger the parameter $\theta_{0}$\,, the closer the flow approaches
each minimal model fixed point and the longer time it spends in its
vicinity, where the roaming model can be described as a combined perturbation
by the relevant and irrelevant fields $\phi_{1,3}$ and $\phi_{3,1}$
\cite{RefZamo,Lassig:1991ab}. Subsequent work resulted in a variety
of generalisations of Zamolodchikov's construction, giving evidence
for further families of RG trajectories interpolating between other
sequences of RG fixed points \cite{RGFlow2,RGFlow3,RGFlow4,RGFlow5}.

A different approach was taken in \cite{RefRoaming}, in which the
authors addressed the calculation of the $c$-function defined by
the $c$-theorem \cite{ZamolodchikovC_function,Cardy_Sum_rule} using
a spectral series in terms of the form factors of the trace of the
stress-energy tensor $\Theta$ \cite{ShGFFZamo,FMS}. The resulting
$c$-function, although not fully identical with the TBA $c$-function,
displays the same plateaux corresponding to the central charges of
the series of conformal minimal models. The central observation of
this work was that the support of the form factor integrals in the
spectral sum comes from specific regions in the multi-dimensional
rapidity space, a phenomenon dubbed relocalisation. These regions
are essentially hypercubes of size $\mathcal{O}(1)$ with centre coordinates
in rapidity space that are integer or half-integer multiples of the
roaming parameter $\theta_{0}$\,. Since $\theta_{0}$ is sent to
infinity in the roaming limit, such hypercubes eventually grow infinitely
separated and become independent cells, from which the form factors
of the trace of stress-energy tensor $\Theta$ in the perturbed minimal
models can be reconstructed.

A logical next step is to attempt the construction of the form factors
of other operators besides $\Theta$ from the roaming trajectories.
For step $k=1$ corresponding to the thermal perturbation of the critical
Ising model, the form factors of the sinh-Gordon field $\phi$ have
been observed to reconstruct the form factors of the magnetisation
operator merely by sending $\theta_{0}\rightarrow\infty$ \cite{ShG-Ising}.
For other steps, however, the limiting procedure is much more difficult,
as the relocalisation patterns studied in \cite{RefRoaming} come
to play a crucial role in the behaviour of the continued sinh-Gordon
form factors.

The step $k=2$ in the roaming limit, corresponding to the massless
flow from the tricritical to the critical Ising model, is particularly
interesting, since besides the presence of the relocalisation patterns,
the form factors of $\Theta\sim\phi_{1,3}$ were explicitly constructed
in the earlier work \cite{MasslessFlow}, together with the form factors
of the order operator $\Phi$ and the disorder operator $\tilde{\Phi}$.
It was verified in \cite{RefRoaming} that the proposed roaming limit
for the form factors of the particular sinh-Gordon operator $\Theta$
indeed matches the results in \cite{MasslessFlow}. Motivated by the
known analytic expressions for the form factors of $\Phi$ and $\tilde{\Phi}$
in the flow between the tricritical and critical Ising models, in
this work we take the first steps towards the construction of the
form factors of other operators besides $\Theta$ in a non trivial
case by showing that they can be obtained as roaming limits of sinh-Gordon
form factors. This leads to a better understanding of the roaming
limits, especially with regard to the role played by CPT symmetry.
In addition, to obtain form factors of $\tilde{\Phi}$ it is necessary
to construct a new solution of the sinh-Gordon form factor bootstrap
which corresponds to a semi-local operator, a.k.a.\ a twist field.

Our paper is organised as follows. In Section \ref{sec:Form-factors-of}
we briefly discuss some general properties of form factors and their
generalisation to massless theories. We show that the form factors
of the order operator $\Phi$ can indeed be obtained from the form
factors of the odd scaling field $:\sinh\frac{g}{2}\phi:$. This construction
necessitates a CPT symmetrisation step, which is argued to be a general
ingredient of the roaming limit procedure. We also show that similarly
to the Ising model, the order operator can be obtaned from the form
factors of the elementary sinh-Gordon field as well. Section \ref{sec:Sinh-Gordon-twist-field}
is devoted to the disorder operator $\tilde{\Phi}$. Due its semi-local
nature, it is first necessary to solve the sinh-Gordon form factor
bootstrap for semi-local operators; we construct the minimal solution
of the recursion up to 6 particles. Using the $\Delta$-theorem sum
rule \cite{delta_theorem} it is shown that the field defined by the
minimal solution is just the off-critical version of the twist operator
known from the free massless scalar field theory. We then show that
in the roaming limit the form factors of the twist field reproduce
those of the disorder operator $\tilde{\Phi}$. Our conclusions are
presented in Section \ref{sec:Conclusions}. To keep the main line
of argument clear, some of the more technical details of the calculations
are relegated to appendices.

\section{Form factors of the order operator \label{sec:Form-factors-of}}

In the following, we discuss two-dimensional relativistically invariant
quantum field theories. Form factors are matrix elements of (semi-)local
operators $O(x,t)$ between the vacuum and asymptotic states, i.e.,

\begin{equation}
F_{\alpha_{1},\ldots,\alpha_{n}}^{O}(\theta_{1},\ldots,\theta_{n})=\langle0|O(0,0)|\theta_{1},\ldots\theta_{n}\rangle_{\alpha_{1},\ldots,\alpha_{n}}.\label{eq:FF}
\end{equation}
In massive theories, the asymptotic states correspond to multi-particle
excitations, with dispersion relation $(E,p)=(m_{\alpha_{i}}\cosh\theta,m_{\alpha_{i}}\sinh\theta)$,
where $\alpha_{i}$ indicates the particle species. In such models,
any multi-particle state can be constructed from vacuum state by means
of the particle creation operators $A_{\alpha_{i}}^{\dagger}(\theta)$
by 
\begin{equation}
|\theta_{1},\theta_{2},...,\theta_{n}\rangle_{\alpha_{1},...,\alpha_{n}}=A_{\alpha_{1}}^{\dagger}(\theta_{1})A_{\alpha_{2}}^{\dagger}(\theta_{2})\ldots.A_{n}^{\dagger}(\theta_{n})|0\rangle\:,\label{eq:basis}
\end{equation}
where the operator $A_{\alpha_{i}}^{\dagger}(\theta)$ creates a particle
of species $\alpha_{i}$ with rapidity $\theta$ and $|0\rangle$
is the vacuum state of the theory. In an integrable QFT with factorised
scattering, the creation and annihilation operators $A_{\alpha_{i}}^{\dagger}(\theta)$
and $A_{\alpha_{i}}(\theta)$ satisfy the Zamolodchikov-Faddeev (ZF)
algebra 
\begin{eqnarray}
A_{\alpha_{i}}^{\dagger}(\theta_{i})A_{\alpha_{j}}^{\dagger}(\theta_{j}) & = & S_{\alpha_{i},\alpha_{j}}(\theta_{i}-\theta_{j})A_{\alpha_{j}}^{\dagger}(\theta_{j})A_{\alpha_{i}}^{\dagger}(\theta_{i})\:,\nonumber \\
A_{\alpha_{i}}(\theta_{i})A_{\alpha_{j}}(\theta_{j}) & = & S_{\alpha_{i},\alpha_{j}}(\theta_{i}-\theta_{j})A_{\alpha_{i}}(\theta_{j})A_{\alpha_{i}}(\theta_{i})\:,\nonumber \\
A_{\alpha_{i}}(\theta_{i})A_{\alpha_{j}}^{\dagger}(\theta_{j}) & = & S_{\alpha_{i},\alpha_{j}}(\theta_{j}-\theta_{i})A_{\alpha_{j}}^{\dagger}(\theta_{j})A_{\alpha_{i}}(\theta_{i})+\delta_{\alpha_{i},\alpha_{j}}2\pi\delta(\theta_{1}-\theta_{2})\boldsymbol{1}\:,\label{eq:ZF}
\end{eqnarray}
where $S_{\alpha_{i},\alpha_{j}}(\theta_{i}-\theta_{j})$ are the
two-particle S-matrices of the theory. However, (\ref{eq:FF}) can
be applied to massless theories as well using the concepts of massless
scattering theory \cite{massless_Smat} with the particle species
labels including also a distinction between left-movers ($L$) and
right-movers ($R$).

The essence of the massless S-matrix approach can be formulated by
taking an appropriate limit of a massive integrable model; for simplicity
we restrict ourselves here to the case of a single self-conjugate
particle since this the case we are mostly interested in. In the massless
model, we have right- and left-moving particles, whose creation operators
can be obtained from the massive creation operators as 
\[
A_{R,L}^{\dagger}(\theta)=\underset{\theta_{0}\rightarrow\infty}{\lim}A^{\dagger}(\theta\pm\beta_{0}/2)\:,
\]
where the mass $m$ in the massive model must be sent to zero so that
the scale $M=me^{\beta_{0}/2}$ stays finite; $M$ is eventually the
cross-over scale along the resulting massless flow. The spectrum of
the right-moving and left-moving particles is then given by 
\begin{eqnarray*}
R & : & p^{0}=p^{1}=\frac{M}{2}e^{\theta}\\
L & : & p^{0}=-p^{1}=\frac{M}{2}e^{-\theta}\:.
\end{eqnarray*}
With the massless creation and annihilation operators the asymptotic
states can be written similarly to the massive cases, and satisfy
an algebra of the form (\ref{eq:FF}) which can be obtained as the
limit of the massive ZF algebra with S-matrices \cite{MasslessFlow}
\begin{eqnarray*}
S_{LL}(\theta) & = & S_{RR}(\theta)=S(\theta)\\
S_{RL}(\theta) & = & \lim_{\beta_{0}\rightarrow\infty}S(\theta+\beta_{0})\\
S_{LR}(\theta) & = & \lim_{\beta_{0}\rightarrow\infty}S(-\theta-\beta_{0})\:.
\end{eqnarray*}
In an integrable QFT the form factors of (semi-)local operators satisfy
the so-called form factor bootstrap equations \cite{bergkarowski,kirillovsmirnov,FFAxioms}

\begin{eqnarray}
F_{\alpha_{1},\ldots,\alpha_{i},\alpha_{i+1},\ldots\alpha_{n}}^{O}(\theta_{1},\ldots\theta_{i},\theta_{i+1},\ldots,\theta_{n}) & = & S_{\alpha_{i},\alpha_{i+1}}(\theta_{i}-\theta_{i+1})\label{eq:FFAxiom1}\\
 &  & \,\times F_{\alpha_{1},\ldots,\alpha_{i+1},\alpha_{i},\ldots\alpha_{n}}^{O}(\theta_{1},\ldots\theta_{i+1},\theta_{i},\ldots,\theta_{n})\nonumber \\
F_{\alpha_{1},\alpha_{2},\ldots\alpha_{n}}^{O}(\theta_{1}+2\pi i,\theta_{2},\ldots,\theta_{n}) & = & e^{2\pi i\gamma}F_{\alpha_{2},\ldots\alpha_{n},\alpha_{1}}^{O}(\theta_{2},\ldots,\theta_{n},\theta_{1})\label{eq:FFAxiom2}\\
-i\underset{\theta'=\theta+i\pi}{\mbox{Res}}F_{\alpha,\alpha,\alpha_{1},\ldots\alpha_{n}}^{O}(\theta',\theta,\theta_{1},\theta_{2},\ldots,\theta_{n}) & = & \left(1-e^{2\pi i\gamma}\prod_{i=1}^{n}S_{\alpha,\alpha_{i}}(\theta-\theta_{i})\right)\label{eq:FFAxiom3}\\
 &  & \,\times F_{\alpha_{1},\ldots\alpha_{n}}^{O}(\theta_{1},\theta_{2},\ldots,\theta_{n})\:.\nonumber 
\end{eqnarray}
The $e^{2\pi i\gamma}$ factor in (\ref{eq:FFAxiom2}) and (\ref{eq:FFAxiom3})
is called the semi-local or mutual locality index of the operator
$O$ with respect to the interpolating field $\phi$ and is defined
via the condition 
\[
O(x,t)\phi(y,t')=e^{2\pi i\gamma}\phi(y,t')O(x,t)
\]
for space-like separated space-time points. Local operators correspond
to $e^{2\pi i\gamma}=1$, while fields with $e^{2\pi i\gamma}\neq1$
are called semi-local.

In addition, relativistic invariance implies

\begin{equation}
F_{\alpha_{1},\ldots,\alpha_{n}}^{O}(\theta_{1}+\Lambda,\ldots,\theta_{n}+\Lambda)=e^{s\Lambda}F_{\alpha_{1},\ldots,\alpha_{n}}^{O}(\theta_{1},\ldots,\theta_{n}),\label{eq:RelInv}
\end{equation}
where $s$ is the Lorentz spin of the operator. As the models considered
in this paper have no bound states, (\ref{eq:FFAxiom1})-(\ref{eq:FFAxiom3})
and (\ref{eq:RelInv}) give all the constraints for form factors of
general (semi-)local operators.

\subsection{Form factors of exponential fields in the sinh-Gordon model}

The sinh-Gordon model is defined by the Hamiltonian 
\begin{eqnarray}
H & = & \int dx\left[\frac{1}{2}\pi^{2}+\frac{1}{2}(\partial_{x}\phi)^{2}+\frac{m_{0}^{2}}{g^{2}}:\cosh g\phi:\right]\:,\label{eq:hamiltonian}\\
 &  & [\phi(t,x),\pi(t,y)]=i\delta(x-y)\nonumber 
\end{eqnarray}
where $m_{0}$ is the classical particle mass and $g$ the coupling
constant. The spectrum of the model consists of multi-particle states
of a single massive bosonic particle with exact mass $m$. The two-particle
$S$-matrix is \cite{AFZ} 
\begin{equation}
S(\theta)=\frac{\sinh(\theta)-(\omega-\omega^{-1})/2}{\sinh(\theta)+(\omega-\omega^{-1})/2}\:,\label{eq:SinhGordonS}
\end{equation}
where $\theta=\theta_{1}-\theta_{2}$ is the relative rapidity of
the particles, 
\begin{equation}
\omega=e^{i\frac{\pi B}{2}},\label{eq:kisomega}
\end{equation}
and $B$ is related to the coupling $g$ by 
\begin{equation}
B\,=\,\frac{2g^{2}}{8\pi+g^{2}}\:.
\end{equation}
For the sinh-Gordon model, solutions of the system (\ref{eq:FFAxiom1})-(\ref{eq:RelInv})
were first constructed in \cite{FMS}. In \cite{SinhGordonFF}, an
important class of form factor solutions related to the exponential
operators 
\begin{equation}
:e^{\kappa g\Phi}:\:\label{eq:ExpOperator}
\end{equation}
was obtained, where $\kappa\in\mathbb{R}$ . The exact vacuum expectation
values of these operators were found in \cite{ExactVEV} and are denoted
here as 
\begin{equation}
G_{\kappa}=H_{0}^{\kappa}=\left\langle 0\right|:e^{\kappa g\Phi}:\left|0\right\rangle .\label{eq:VEV}
\end{equation}
Introducing the notation 
\begin{equation}
O^{\kappa}=G_{\kappa}^{-1}:e^{\kappa g\Phi}:\,,\label{eq:Okappa}
\end{equation}
the form factors of normalised exponential operators take the following
form \cite{SinhGordonFF} 
\begin{equation}
F_{n}^{\kappa}\left(\theta_{1},\dots,\theta_{n}\right)=\left\langle 0\right|O^{\kappa}|\theta_{1},\dots,\theta_{n}\rangle=\left(\frac{4\sin\frac{\pi B}{2}}{\mathcal{N}}\right)^{n/2}\frac{Q_{n}^{\kappa}\left(x_{1},\dots,x_{n}\right)}{\prod_{i<j}\left(x_{i}+x_{j}\right)}\prod_{i<j}f_{min}\left(\theta_{i}-\theta_{j}\right)\,,\label{eq:FFParametrizationShGLocal}
\end{equation}
where $x_{i}=e^{\theta_{i}}$. In fact, apart from the overall normalisation,
(\ref{eq:FFParametrizationShGLocal}) gives the most general parametrisation
for the form factors of local operators \cite{ShGFFZamo,FMS}. The
minimal form factor $f_{min}$ is

\begin{eqnarray}
f_{min}\left(\theta,B\right) & = & \mathcal{N}\:\tilde{f}_{min}\left(\theta,B\right),\label{eq:FminShG}\\
 &  & \tilde{f}_{min}\left(\theta,B\right)=\exp\left[8\int_{0}^{\infty}\frac{\mathrm{d}t}{t}\frac{\sinh\frac{Bt}{4}\sinh\frac{t}{2}\left(1-\frac{B}{2}\right)\sinh\frac{t}{2}}{\sinh^{2}t}\sin^{2}\left[\frac{t(i\pi-\theta)}{2\pi}\right]\right]\:,\nonumber 
\end{eqnarray}
which is a complex valued meromorphic function without singularities
for real rapidities. This function tends to unity for large rapidities,
i.e., $\underset{\theta\rightarrow\pm\infty}{\lim}f_{min}(\theta)=1$,
so long as the normalisation is chosen to be

\begin{equation}
\mathcal{N}=f_{min}\left(i\pi,B\right)=\exp\left[-4\int_{0}^{\infty}\frac{\mathrm{d}t}{t}\frac{\sinh\frac{Bt}{4}\sinh\frac{t}{2}\left(1-\frac{B}{2}\right)\sinh\frac{t}{2}}{\sinh^{2}t}\right]\:.\label{eq:FminNormShG}
\end{equation}

The functions $Q_{n}$ appearing in (\ref{eq:FFParametrizationShGLocal})
are entire functions and completely symmetric in the variables $x_{i}$;
for an operator with a power-like short distance singularity they
can only grow exponentially for large values of the rapidities and
are therefore restricted to be polynomials \cite{FFgrowth}. From
now on we call these functions \emph{form factor polynomials}; their
growth at infinity is related to the ultraviolet scaling dimension
of the operator, and solutions with the lowest possible growth at
infinity, called minimal solutions, correspond to operators with the
lowest possible conformal dimensions. Since the form (\ref{eq:FFParametrizationShGLocal})
of the form factor is completely general, the dependence on the specific
operator is carried by the form factor polynomials $Q_{n}$.

For the exponential operators discussed here, the polynomial part
can be written as 
\begin{equation}
Q_{n}^{\kappa}=\det M_{ij}(\kappa)\:,\label{eq:Q_nDet}
\end{equation}
where $M$ is an $(n-1)\times(n-1)$ matrix with elements

\begin{equation}
M_{ij}(\kappa)=\left[\kappa+i-j\right]\sigma_{2i-j}^{(n)}\:,\label{eq:Q_nMatrix}
\end{equation}
where 
\begin{equation}
\left[n\right]=\frac{\sin n\frac{\pi B}{2}}{\sin\frac{\pi B}{2}}\:,\label{eq:Bracket}
\end{equation}
and $\sigma_{i}^{(n)}$ denotes the $i$th symmetric polynomial of
$n$ variables $x_{1},\ldots,x_{n}$ defined by the generating function
\[
\prod_{i=1}^{n}(x+x_{i})=\sum_{k=0}^{n}x^{n-k}\sigma_{k}^{(n)}(x_{1},\dots,x_{n})\:.
\]
The upper index $(n)$ will be omitted in cases where this causes
no confusion.

As the exponential operators are spinless, the total degree of the
polynomials (\ref{eq:Q_nDet}) must be the same as that of the denominator
in (\ref{eq:FFParametrizationShGLocal}); since the total degree of
$Q_{n}^{\kappa}$ is $n(n-1)/2$ this is indeed satisfied. The partial
degree of the polynomials is at most $n-1$, which ensures that $\underset{\theta_{i}\rightarrow\infty}{\lim}F_{n}^{\kappa}(\theta_{1},\ldots,\theta_{n})$
is bounded by a constant. A further important property of the form
factors of exponential operators is clustering (also called asymptotic
factorisation) \cite{delta_theorem}:

\begin{equation}
\underset{\Lambda\rightarrow\infty}{\lim}F_{r+l}^{O}(\theta_{1}+\Lambda,\ldots,\theta_{r}+\Lambda,\theta_{r+1},\ldots\theta_{r+l})=\frac{1}{\langle O\rangle}F_{r}^{O}(\theta_{1},\ldots,\theta_{r})F_{l}^{O}(\theta_{r+1},\ldots\theta_{r+l})\:.\label{eq:ClusterProp}
\end{equation}
When applied to form factors of normalised exponential operators this
can be written as

\begin{equation}
\underset{\Lambda\rightarrow\infty}{\lim}F_{r+l}^{\kappa}(\theta_{1}+\Lambda,\ldots,\theta_{r}+\Lambda,\theta_{r+1},\ldots\theta_{r+l})=F_{r}^{\kappa}(\theta_{1},\ldots,\theta_{r})F_{l}^{\kappa}(\theta_{r+1},\ldots\theta_{r+l})\:.\label{eq:ClusterPropExpOp}
\end{equation}
Finally, we remark that the form factors of the elementary field $\phi$
in the sinh-Gordon model are proportional to $\frac{dF_{n}^{\kappa}}{d\kappa}|_{\kappa=0}$;
in particular, the polynomial part is just $Q_{n}^{0}=\det M_{ij}(0)$.
The form factors of $\phi$ are only non-zero when $n$ is odd.

\subsection{Construction of the order operator for the massless flow}

Similarly to the sinh-Gordon form factors (\ref{eq:FFParametrizationShGLocal}),
the form factors of the massless flow can be written as \cite{MasslessFlow}

\begin{equation}
\begin{array}{cc}
F_{r,l}(\theta_{1},\ldots,\theta_{r},\theta'_{1},\ldots,\theta'_{l})= & H_{r,l}Q_{r,l}(x_{1},\ldots,x_{r},y_{1}\ldots y_{l})\underset{1\leqslant i<j\leqslant r}{\prod}\frac{f_{RR}(\theta_{i}-\theta_{j})}{x_{i}+x_{j}}\\
 & \times\stackrel[i=1]{r}{\prod}\stackrel[j=1]{l}{\prod}f_{RL}(\theta_{i}-\theta'_{j})\underset{1\leqslant i<j\leqslant l}{\prod}\frac{f_{LL}(\theta'_{i}-\theta'_{j})}{y_{i}+y_{j}}\:.
\end{array}\label{eq:FFParametrizationFlow}
\end{equation}
In the above formula we omitted the reference to the specific operator;
indices $r$ and $l$ refer to the number of right/left-moving particles,
respectively. $H_{r,l}$ is a suitably chosen normalisation constant,
and the minimal two-particle form factors are 
\begin{equation}
f_{RR}(\theta)=f_{LL}(\theta)=\sinh(\theta/2)\label{eq:fRR}
\end{equation}
and 
\begin{equation}
f_{RL}(\theta)=\exp\left(\frac{\theta}{4}-\int_{0}^{\infty}\frac{dt}{t}\frac{\sin^{2}(\frac{i\pi-\theta}{2\pi}t)}{\sinh(t)\cosh(t/2)}\right)\:.\label{eq:fRL}
\end{equation}
The $Q_{r,l}$ are functions of the variables $x_{i}=e^{\theta_{i}}$
and $y_{i}=e^{-\theta'_{i}}$; for local operators, $Q_{r,l}$ is
an entire function symmetric separately in the variables $x_{i}$
and in $y_{j}$, while for semi-local operators ($\gamma=1/2$), it
contains also the factor 
\[
\sqrt{\stackrel[i=1]{r}{\prod}x_{i}\stackrel[j=1]{l}{\prod}y_{j}}\,.
\]
For the order and disorder operators $\Phi$ and $\tilde{\Phi}$ the
functions $Q_{r,l}$ were calculated in \cite{MasslessFlow} based
on the form factor bootstrap, however, in our work we slightly modify
the normalisations $H_{r,l}$ for the form factors as described in
Appendix \ref{sub:Appendix-A-MasslessFF}.

To construct the order operator, one needs to find a suitable candidate.
Note that the order operator of the massless flow is odd and its operator
product expansion with itself contains the trace of the energy-momentum
tensor, which is the perturbing field. The most natural candidate
for a starting point is a scaling field in the sinh-Gordon model with
these properties, which is given by 
\[
\tilde{O}^{1/2}=G_{1/2}^{-1}:\sinh\frac{1}{2}g\phi:
\]
Its form factors are identical with those of either $O^{1/2}$ or
$-O^{-1/2}$ whenever the number of particles is odd, but vanish when
their number is even. The normalization factor is chosen as in \eqref{eq:Okappa}.

The roaming limit procedure consists of the following steps. First
the coupling-dependent parameter $B$ is continued as \cite{RefZamo}
\[
B\rightarrow B(\theta_{0}):=1+i\frac{2}{\pi}\theta_{0}\:.
\]
In the form factors of the sinh-Gordon fields $\phi$ and $\Theta$,
making this substitution and sending $\theta_{0}\rightarrow\pm\infty$
is sufficient to recover the form factors of operators $\sigma$ and
$\epsilon$ for the $k=1$ step corresponding to the thermal perturbation
of the critical Ising model, in which $\sigma$ is the magnetization
operator and $\epsilon$ is the energy operator. For other steps,
however, it is necessary to make a second step, to identify the relevant
regions of rapidity space which contribute to spectral sums when taking
$\theta_{0}$ to infinity. Following \cite{RefRoaming}, we introduce
the concept of cells which are labeled by a sequence of monotonically
increasing integers $p_{1}<p_{2}...<p_{l}$ as follows

\begin{equation}
\mathcal{C}=\begin{array}{c}
[\underbrace{p_{l},p_{l},...,p_{l}}_{r_{p_{l}}},\underbrace{p_{l-1},p_{l-1},...,p_{l-1}}_{r_{p_{l-1}}},...,\underbrace{p_{1},p_{1},...,p_{1}}_{r_{p_{1}}}]\end{array}\:,\label{eq:Cell}
\end{equation}
where $r_{p_{1}}...,r_{p_{l}}$ indicate the number of occurrences
of each number $p_{i}$\,, or equivalently the length of each block
of integers. Such a sequence specifies the appropriate limit of the
form factor to be taken, as specified in \cite{RefRoaming} for the
trace of the stress energy tensor $\Theta$. In \cite{RefRoaming}
it was shown that the dominant contribution to the spectral sum comes
from regions in the rapidity space that are hypercubes of width $\mathcal{O}(1)$
and whose centre coordinates can be written as a set of some integers
multiplied with $\theta_{0}/2$. In addition, the only contributions
surviving in the roaming limit are those of these dominant regions.
The cell notation (\ref{eq:Cell}) introduced above identifies each
of these regions with their centre coordinates in units of $\theta_{0}/2$.
Denoting the $i^{{\rm th}}$ element of a cell by $\mathcal{C}_{i}$\,,
the roaming limit for $\Theta$ can be written as

\[
F_{\mathcal{C}}^{\Theta}(\theta_{1},...,\theta_{n})=\underset{\theta_{0}\rightarrow\infty}{\lim}\mathcal{N}e^{-\alpha\theta_{0}}F_{n}^{\Theta}(\theta_{1}+\mathcal{C}_{1}\theta_{0}/2,...,\theta_{n}+\mathcal{C}_{n}\theta_{0}/2,B(\pm\theta_{0}))
\]
with $\alpha$ chosen so that the limit is finite and \emph{$\mathcal{N}$}
introduced to ensure proper normalisation of the minimal model $\Theta$
operator. For the thermally-perturbed Ising model, the only relevant
cell is {[}0,0{]} corresponding to no rapidity shifts; for $k=2$,
i.e., the massless flow from the tricritical to critical Ising, the
only relevant cells are of the type

\begin{equation}
[\underbrace{1,1..,1,}_{r}\underbrace{-1,-1,...,-1}_{l}]\label{eq:CellStep2}
\end{equation}
with $r$, $l$ even, so the minimal model form factors of $\Theta$
can be written as

\begin{equation}
F_{r,l}^{\Theta}=\underset{\theta_{0}\rightarrow\infty}{\lim}\mathcal{N}e^{-2\theta_{0}}F_{r+l}^{\Theta}(\theta_{1}+\theta_{0}/2,...,\theta_{r}+\theta_{0}/2,\theta_{r+1}-\theta_{0}/2,...,\theta_{r+l}-\theta_{0},B(\pm\theta_{0}))\label{eq:Thetalim}
\end{equation}
in terms of the sinh-Gordon form factors and the phase factor $\mathcal{N}$
accounts for the difference between the proper normalisation of the
two-particle form factor. We see that the cell structure accounts
for the structure of the massless form factor through the separation
of right- and left-moving particles. For higher flows the procedure
described in \cite{RefRoaming} also generates so-called ``magnonic''
degrees of freedom that correspond to the internal structure of the
kink excitations in the massless scattering theory. We omit these
details as they are not needed in the sequel.

For the operator $\Theta$ a natural identification of the dominant
integration regions \cite{RefRoaming} is given by the $c$-theorem
spectral sum, which can be formulated as \cite{ZamolodchikovC_function,Cardy_Sum_rule}

\begin{equation}
c(R)=c(\infty)+\frac{3}{2}\int_{R}^{\infty}\mathrm{d}R'(R')^{3}\langle\Theta(R')\Theta(0)\rangle\:,\label{eq:CtheoremSumRule}
\end{equation}
in which $R$ is a length-scale. For the massive sinh-Gordon theory
with $c(\infty)=0,$ using the form factor expansion of the correlation
function $\langle\Theta(R')\Theta(0)\rangle$, (\ref{eq:CtheoremSumRule})
can be rewritten with the form factors of $\Theta$ as

\begin{equation}
c(r)=3\stackrel[n=0]{\infty}{\sum}\int_{\mathbb{R^{n}}}\frac{\mathrm{d}\theta_{1}...\mathrm{d}\theta_{n}}{(2\pi)^{n}n!}\frac{6+6rE_{n}+3r^{2}E_{n}^{2}+r^{3}E_{n}^{3}}{2E_{n}^{4}}|m^{-2}F^{\Theta}(\theta_{1},...\theta_{n})|e^{-rE}\:,\label{eq:CTheoremSumRuleFF}
\end{equation}
where $m$ is the physical mass, $r=mR$ and $E_{n}=\stackrel[i=1]{n}{\sum}\cosh\theta_{i}$.
As shown in \cite{RefRoaming}, it is possible to carry out the roaming
limit within the $c$-theorem sum rule; we briefly summarise the relevant
arguments here.

First, the overall asymptotic behaviour of the form factors of $\Theta$
was found to be $e^{N\theta_{0}}$, with $N$ denoting the number
of blocks in the cells, as long as all the blocks consist of an even
number of members and the difference between the members of neighbouring
blocks is 2. If these criteria are not fulfilled, the asymptotics
is always given by $e^{\omega\theta_{0}}$ with $\omega<N$. The number
of blocks $N$ and the members of the cells are constrained on one
hand by the number of variables of the form factors, and on the other
hand by the exponential factor and the energy denominator entering
the $c$-theorem sum rule integrands. To see the $k^{{\rm th}}$ step
in the roaming limit, the distance $r$ must scale as $r=\tilde{r}e^{-(k-1)\frac{\theta_{0}}{2}}$,
where $\tilde{r}$ is finite, but due to the exponential factor in
(\ref{eq:CTheoremSumRuleFF}), $|\mathcal{C}_{i}|\leq k-1$. The contribution
of cells with $|\mathcal{C}_{i}|\geq k-1$ are suppressed by a double
exponential way. In addition to this, to minimise the energy term
in the denominator one finds that 
\[
\min_{i}(\mathcal{C}_{i})=-\max_{i}(\mathcal{C}_{i})\,.
\]
These considerations leads to the following pattern for the dominant
cells

\begin{equation}
[p_{l},p_{l},...p_{1},p_{1}]:l\geq k,-p_{1}=p_{l}=k-1,p_{k}-p_{k-1}=2\text{ or }0
\end{equation}
for the $k^{{\rm th}}$ step of the staircase; for $k=2$ this reproduces
(\ref{eq:CellStep2}). This yields the limiting procedure (\ref{eq:Thetalim})
for $\Theta$ with the normalisation of the operator fixed by the
$c$-theorem.

For sinh-Gordon operators with form factors of even particle number,
the $c$-theorem could be replaced by the $\Delta$-theorem \cite{delta_theorem}
to define the dominant integration cells and hence the eventual limit
of the form factors. If it converges, the $\Delta$-theorem states
that if at some length scale $R$ the theory can be described by a
CFT, then the difference of the conformal weight of an operator $O$
and its conformal weight in the IR limit can be calculated as

\begin{eqnarray}
D(R)-\Delta^{IR} & = & -\frac{1}{4\pi\left\langle O\right\rangle }\int_{x^{2}>R}\mathrm{d^{2}}x\langle\Theta(x)O(0)\rangle\:.\label{eq:DeltaTheoreM1}
\end{eqnarray}
Similarly to the $c$-theorem, (\ref{eq:DeltaTheoreM1}) has a spectral
representation 
\begin{equation}
D(r)-\Delta^{IR}=-\frac{1}{2\left\langle O\right\rangle }\sum_{n=1}^{\infty}\int\frac{\mathrm{d}\theta_{1}...\mathrm{d}\theta_{n}}{(2\pi)^{n}n!}\frac{e^{-rE_{n}}(1+E_{n}r)}{E_{n}^{2}}m^{-2}F^{\Theta}\left(\theta_{1},\dots,\theta_{n}\right)F^{O}\left(\theta_{n},\dots,\theta_{1}\right)\:.\label{eq:DeltaTheoreM2}
\end{equation}
However, operators $\Phi$ and $\tilde{O}^{1/2}$ have non vanishing
form factors only when $r+l$ is odd, and for these the $\Delta$-theorem
is eventually vacuous, since $\Theta$ has only even form factors
and also $\left\langle \Phi\right\rangle =0$. But the form factor
expansion of the two point correlation function 
\begin{equation}
\langle\tilde{O}^{1/2}(x)\tilde{O}^{1/2}(0)\rangle=\sum_{n=1}^{\infty}\int\frac{\mathrm{d}\theta_{1}}{2\pi}\dots\frac{\mathrm{d}\theta_{2n+1}}{2\pi}\frac{1}{(2n+1)!}|F^{1/2}(\theta_{1},..\theta_{2n+1})|^{2}e^{-rE_{2n+1}}\:\label{eq:O-O2PointFunction}
\end{equation}
can be used to find the dominant cells for the roaming limit of the
form factors of $\Phi$. Note that due to the definition of $\tilde{O}^{1/2}$
only form factors of $O^{1/2}$ with odd number of particles appear
in (\ref{eq:O-O2PointFunction}).

The only essential difference between the structures of (\ref{eq:CTheoremSumRuleFF})
and (\ref{eq:O-O2PointFunction}) is the absence of the energy denominator
in the latter, which means that the condition 
\[
\min_{i}(\mathcal{C}_{i})=-\max_{i}(\mathcal{C}_{i})
\]
is relaxed. Following the power counting method used in \cite{RefRoaming},
the asymptotic behaviour of the form factors of $O^{1/2}$ can be
obtained as

\begin{equation}
{e^{\frac{1}{4}N_{odd}\theta_{0}+\theta_{0}/2}\:,}\label{eq:Nodd}
\end{equation}
where $N_{odd}$ is the number of blocks with an odd number of members.
For $k=2,$ the exponential part in (\ref{eq:O-O2PointFunction})
allows cells with $\mathcal{C}_{i}=\pm$1 only, and therefore the
dominant cells for correlation function of $\tilde{O}$ are again
(\ref{eq:CellStep2}), but now $r+l$ must be odd and at least 1.

Having identified the dominant cells for the spectral sum of the roaming
two-point function for the $k=2$ step, we can finally perform the
limit at the level of the form factors. Consider the following limit
for odd $n=r+l$ 
\begin{equation}
\underset{\theta_{0}\rightarrow\infty}{\lim}F_{n}^{1/2}(\theta_{1}+\theta_{0},\ldots,\theta_{r}+\theta_{0},\theta'_{1},\ldots,\theta'_{l},B(\pm\theta_{0}))\:,\label{eq:FFLimitFormulation}
\end{equation}
where we used Lorentz invariance to rearrange the rapidity shifts.
Note that due to the asymptotic behaviour (\ref{eq:Nodd}) it is necessary
to renormalise the form factor by an appropriate power of $e^{\theta_{0}}$
in order to obtain finite quantities. However, this power is independent
of the particle number, therefore it makes sense to take the limit
of the form factors similarly to the case of $\Theta$.

Let us analyse the limit now in detail. The limit of the unnormalised
minimal form factors $\tilde{f}_{min}$ in the sinh-Gordon model is
\begin{equation}
\begin{array}{cc}
\underset{\theta_{0}\rightarrow\infty}{\lim}\tilde{f}_{min}(\theta+k\theta_{0},B(\pm\theta_{0}))= & \begin{cases}
-i\sinh\theta/2 & k=0\\
e^{\theta_{0}/2}\frac{1-i}{4}2^{^{1/4}}e^{-K/\pi}\exp\left(\frac{\theta}{4}-\int_{0}^{\infty}\frac{dt}{t}\frac{\sin^{2}(\frac{i\pi-\theta}{2\pi}t)}{\sinh(t)\cosh(t/2)}\right) & k=1\:,
\end{cases}\end{array}\label{eq:fminLim}
\end{equation}
where $K$ is Catalan's constant. (\ref{eq:fminLim}) reproduces the
minimal form factors of the flow, $f_{RR}=f_{LL}=\sinh\theta/2$ and
(\ref{eq:fRL}) as obtained in \cite{MasslessFlow}, up to some constant
normalization factor. For a sinh-Gordon form factor with $n$ rapidities,
separated into $r$ right-moving and $l$ left-moving rapidities in
the limiting procedure, the limit of the form factor and the minimal
form factor normalizations including also the $(rl)^{{\rm th}}$ power
of the term $e^{\theta_{0}/2}$ from (\ref{eq:fminLim}) in the leading
order of $\mathcal{O}(e^{\theta_{0}})$ is

\begin{equation}
e^{rl\theta_{0}/2}e^{\theta_{0}(n(4-n)/4)}\cdot(-i)^{r(r-1)/2+l(l-1)/2}\left(\frac{1-i}{4}2^{^{1/4}}e^{-K/\pi}\right)^{lr}2^{n(n-1)/2}\:.\label{eq:NormalizationLim}
\end{equation}
For the limit of the kinematic pole denominator we obtain 
\begin{equation}
\stackrel[1\leq i<j\leq n]{}{\prod}(x_{i}+x_{j})\rightarrow e^{\theta_{0}(r(r-1)/2+rl)}\stackrel[1\leq i<j\leq r]{}{\prod}(x_{i}+x_{j})\stackrel[1\leq n<m\leq l]{}{\prod}\frac{y_{n}+y_{m}}{y_{n}y_{m}}\stackrel[p=1]{r}{\prod}x_{p}^{l}\:,\label{eq:DenomLimitMainText}
\end{equation}
where $x_{i}=e^{\theta_{i}}$ ($i=1,...,r)$ and $y_{i}=e^{-\theta'_{j}}$
$(j=1,...,l$). Note that in order to match the structure of the massless
form factor (\ref{eq:FFParametrizationFlow}), the factors
\[
\stackrel[1\leq n<m\leq l]{}{\prod}y_{n}y_{m}\stackrel[p=1]{r}{\prod}x_{p}^{-l}
\]
resulting from the limit of denominator must be combined with the
limit of the $Q_{n}$ polynomials to give the functions $\tilde{Q}_{r,l}$,
generally resulting in a rational function of symmetric polynomials.

For the roaming limit of the $Q_{n}$ polynomials we present some
explicit results. For $n=1$ the limit is simply 
\[
e^{3\theta_{0}/4}\,,
\]
where the $\pm$ corresponds to taking the roaming limit with either
the $B(\theta_{0})$ or the $B(-\theta_{0})$ substitution.

For $n=3$, the $\tilde{Q}_{r,l}$ functions obtained from the roaming
limit together with the various powers of $e^{\theta_{0}}$ extracted
from the different parts of the form factor function are given in
the table below:

\begin{center}
\begin{tabular}{|c|c|c|c|c|c|}
\hline 
$n=3$  & $\tilde{Q}_{r,l}$  & Normalisation  & $Q_{n}$  & Denominator  & Overall scaling\tabularnewline
\hline 
\hline 
$r=0$  & $y_{1}y_{2}y_{3}$  & $e^{3\theta_{0}/4}$  & 1 & 1  & $e^{3\theta_{0}/4}$\tabularnewline
\hline 
$r=1$  & $\frac{1}{x_{1}}\pm i(y_{1}+y_{2})$  & $e^{7\theta_{0}/4}$  & $e^{\theta_{0}}$  & $e^{-2\theta_{0}}$  & $e^{3\theta_{0}/4}$\tabularnewline
\hline 
$r=2$  & $\frac{1}{y_{1}}\pm i(x_{1}+x_{2}$  & $e^{7\theta_{0}/4}$  & $e^{2\theta_{0}}$  & $e^{-3\theta_{0}}$  & $e^{3\theta_{0}/4}$\tabularnewline
\hline 
$r=3$  & $x_{1}x_{2}x_{3}$  & $e^{3\theta_{0}/4}$  & $e^{3\theta_{0}}$  & $e^{-3\theta_{0}}$  & $e^{3\theta_{0}/4}$\tabularnewline
\hline 
\end{tabular}
\par\end{center}

\noindent For $n=5$ the corresponding tables are presented in Appendix
\ref{sub:Appendix-C-Tables}.

We see that the overall scaling of the form factors is indeed independent
of the number of particles and is equal to 
\[
e^{3\theta_{0}/4}\,,
\]
which can be proven using the power counting approach of \cite{RefRoaming}.
This allows a general renormalisation of the form factors that is
independent of the number of particles to be applied before taking
the roaming limit. We also include a finite renormalisation so that
the roaming limits of the one-particle form factors satisfy $\langle0|\Phi|\theta\rangle_{R/L}=1.$
This is a natural normalisation since we expect $\langle0|\Phi|\theta\rangle_{R}=\langle0|\Phi|\theta\rangle_{L}$
due to parity invariance.

Comparing the resulting form factor polynomials with the known exact
expressions for the minimal model form factors \cite{MasslessFlow}
(cf.\ also Appendix \ref{sub:Appendix-A-MasslessFF}) shows that
the form factor polynomials obtained from the roaming limit contain
an additional imaginary part which changes sign when switching between
the two possible roaming limits corresponding to $B(+\theta_{0})$
and $B(-\theta_{0})$.

The solution of this discrepancy lies in the observation that both
the sinh-Gordon model and the massless flow are CPT invariant, since
they are described by Hermitian relativistic invariant actions. However,
the naive roaming continuation breaks hermiticity, and CPT takes $B(+\theta_{0})$
into $B(-\theta_{0})$ and vice versa. On the other hand, both roaming
continuations yield a solution for the form factor bootstrap for the
massless form factor bootstrap since the sinh-Gordon $S$-matrix (\ref{eq:SinhGordonS})
is mapped into the massless S-matrix. As a result, any solution for
the sinh-Gordon form factor bootstrap is taken to a solution for the
form factor bootstrap of the massless flow. However, the form factor
bootstrap is linear, and the correspondence between its solutions
and local operators is not necessarily trivial.

As a result of the above considerations we propose a further CPT symmetrisation
step before the roaming limit and to consider the expression 
\begin{equation}
\begin{array}{ccc}
 & \underset{\theta_{0}\rightarrow\infty}{\lim} & \Bigg(\frac{F_{n}^{1/2}(\theta_{1}+\theta_{0},\ldots,\theta_{r}+\theta_{0},\theta'_{1},\ldots,\theta'_{l},B(+\theta_{0}))}{2F_{1}^{1/2}(0,B(+\theta_{0}))}\\
 &  & +\frac{F_{n}^{1/2}(\theta_{1}+\theta_{0},\ldots,\theta_{r}+\theta_{0},\theta'_{1},\ldots,\theta'_{l},B(-\theta_{0}))}{2F_{1}^{1/2}(0,B(-\theta_{0}))}\Bigg)
\end{array}\label{eq:FFLimitFormulation1}
\end{equation}
for odd $n$. This results in the following result for the CPT symmetrised
$Q_{r,l}$ functions as obtained from the roaming limit, which is
equal to the known exact expressions (\cite{MasslessFlow}, cf.\ also
Appendix \ref{sub:Appendix-A-MasslessFF})

\begin{equation}
\begin{array}{ll}
Q_{r,0}=\rho_{r}^{(r-1)/2} & \text{for odd }r\\
Q_{r,1}=\frac{\rho_{r}^{r/2-1}}{\lambda_{1}^{r/2}} & \text{for even }r\\
Q_{r,2}=\rho_{r}^{(r-3)/2}\stackrel[k=0]{r}{\sum'}\rho_{k}\lambda_{2}^{(k-r+1)/2} & \text{for odd }r\\
Q_{r,3}=\frac{\rho_{r}^{r/2-2}}{\lambda_{3}^{r/2-1}}\stackrel[k=0]{r}{\sum'}\rho_{k}\lambda_{2}^{k/2} & \text{for even }r\:,
\end{array}\label{eq:QOrderRight}
\end{equation}
where $\rho_{k}$ denotes the $k$th symmetric polynomial of variables
$x_{1},...,x_{r}$, $\lambda_{k}$ denotes the $k$th symmetric polynomial
of variables $y_{1},...,y_{l}$ and the primed sum means summation
on even indices.

The remaining task is to check the normalisation factors (\cite{MasslessFlow},
cf.\ also Appendix \ref{sub:Appendix-A-MasslessFF})

\begin{equation}
H_{r,l}=2^{r(r-1)/2+l(l-1)/2}\gamma^{-rl/2}i^{(rl+r+l-1)/2}\:,\label{eq:FFOrderExactNorm}
\end{equation}
where $\gamma=\sqrt{2}e^{2K/\pi}$. It is not difficult to see that
this is identical with (\ref{eq:NormalizationLim}) when the latter
is rescaled with the appropriate power of $e^{\theta_{0}}$. Therefore,
the final form of the limit that gives the normalised form factors
of the order operator is

\begin{equation}
\begin{array}{ccc}
F_{r,l}^{\Phi} & =\underset{\theta_{0}\rightarrow\infty}{\lim} & \Bigg(\frac{F_{n}^{1/2}(\theta_{1}+\theta_{0},\ldots,\theta_{r}+\theta_{0},\theta'_{1},\ldots,\theta'_{l},B(+\theta_{0}))}{2F_{1}^{1/2}(B(+\theta_{0}))}\\
 &  & +\frac{F_{n}^{1/2}(\theta_{1}+\theta_{0},\ldots,\theta_{r}+\theta_{0},\theta'_{1},\ldots,\theta'_{l},B(-\theta_{0}))}{2F_{1}^{1/2}(B(-\theta_{0}))}\Bigg)
\end{array}\label{eq:FFLimitFormulationFinal}
\end{equation}
for $n=r+l$ odd. Note that since the solution for the sinh-Gordon
form factors is known for any number of particles in a closed form,
our result goes beyond the one obtained in the massless form factor
bootstrap since it gives the form factors in a closed form for any
number of particles instead of a recursive construction.

In the above derivation, the form factors of the order operator for
the massless flow were reconstructed from the form factors of $\tilde{O}^{1/2}$,
which is proportional to $:\sinh\frac{1}{2}g\phi:$ . In addition,
it is easy to show that for the first step of the staircase, the form
factors of the order operator in the thermal perturbed Ising model
can also be obtained from this field, as for odd $n$, following the
construction presented in \cite{ShG-Ising} that consists of simply
sending $\theta_{0}$ to $\infty$, one has

\begin{equation}
\underset{\theta_{0}\rightarrow\infty}{\lim}\frac{F_{n}^{1/2}(\theta_{1},\ldots,\theta_{n},B(\pm\theta_{0}))}{F_{1}^{1/2}(B(\pm\theta_{0}))}=i^{(n-1)/2}\stackrel[1\leq i<j\leq n]{}{\prod}\tanh(\frac{\theta_{i}-\theta{}_{j}}{2})\:.\label{eq:LimitOrderFinalPerturbedIsing}
\end{equation}
In \cite{ShG-Ising}, the form factors of the order operator in the
thermal perturbed Ising model were reconstructed from the form factors
of the sinh-Gordon field $\phi$ as

\begin{equation}
\underset{\theta_{0}\rightarrow\infty}{\lim}\frac{F_{n}^{\phi}(\theta_{1},\ldots,\theta_{n},B(\pm\theta_{0}))}{F_{1}^{\phi}(B(\pm\theta_{0}))}=i^{(n-1)/2}\stackrel[1\leq i<j\leq n]{}{\prod}\tanh(\frac{\theta_{i}-\theta{}_{j}}{2})\:,\label{eq:LimitOrderFinalPerturbedIsing2}
\end{equation}
note, however, that $\phi$ is not a scaling field. 

It turns out that this works for the case $k=2$ as well: the dominant
cells for $F_{n}^{\phi}$ are identical to those of $F_{n}^{1/2}$
for odd $n$, and the roaming limits of the polynomials $Q_{n}^{0}$
are proportional to the real parts of the limits of $Q_{n}^{1/2}$
. Although their scaling with $\theta_{0}$ is different, the difference
in the scaling exponent is the same for any choice $n=r+l$ provided
$n$ is odd. Therefore one can also write

\begin{equation}
\begin{array}{ccc}
F_{r,l}^{\Phi} & =\underset{\theta_{0}\rightarrow\infty}{\lim} & \frac{F_{n}^{\phi}(\theta_{1}+\theta_{0},\ldots,\theta_{r}+\theta_{0},\theta'_{1},\ldots,\theta'_{l},B(\pm\theta_{0}))}{F_{1}^{\phi}(B(\pm\theta_{0}))}\end{array}\:.\label{eq:FFLimitFormulationFinal2}
\end{equation}
For $\phi$ there is no need of CPT symmetrisation; note that the
roaming limit for the trace of the stress-energy tensor \eqref{eq:Thetalim},
together with the limit for the step $k=1$ order operator and for
\eqref{eq:FFLimitFormulationFinal2} is independent of the sign choice
in $B(\pm\theta_{0})$, therefore the CPT symmetrisation prescription
can be trivially extended to cover that case as well. The fact that
the same operator can be obtained as the roaming limit of more than
one sinh-Gordon field is not surprising. In fact, many more solutions
to the form factor equations of the massless flow can be obtained
by taking limits of other sinh-Gordon fields. However, identifying
them with concrete operators in the perturbed minimal model is far
from trivial. For the case of the order field the existence of the
known exact solution was helpful, since unfortunately in this case
the $\Delta$-theorem sum rule \eqref{eq:DeltaTheoreM2} is vacuous.

\section{Sinh-Gordon twist field and the disorder operator \label{sec:Sinh-Gordon-twist-field}}

\subsection{Semi-local form factors in the sinh-Gordon model}

In contrast to the order field $\Phi$, the disorder field $\tilde{\Phi}$
is a semi-local operator with mutual locality index $e^{2\pi i\gamma}=-1$
\cite{MasslessFlow}. Therefore it is necessary to construct the form
factors of some semi-local operator at the sinh-Gordon level, since
the roaming limit cannot change the mutual locality index.

The form factor equations for a semi-local field in the sinh-Gordon
model that are to be satisfied by the semi-local sinh-Gordon form
factors are the following 
\begin{eqnarray}
F_{n}^{O}(\theta_{1},\ldots\theta_{i},\theta_{i+1},\ldots,\theta_{n}) & = & S(\theta_{i}-\theta_{i+1})F_{n}^{O}(\theta_{1},\ldots\theta_{i+1},\theta_{i},\ldots,\theta_{n})\label{eq:FFAxiom1SemiLocal}\\
F_{n}^{O}(\theta_{1}+2\pi i,\theta_{2},\ldots,\theta_{n}) & = & -F_{n}^{O}(\theta_{2},\ldots,\theta_{n},\theta_{1})\label{eq:FFAxiom2SemiLocal}\\
-i\underset{\theta'=\theta+i\pi}{\mbox{Res}}F_{n}^{O}(\theta',\theta,\theta_{1},\theta_{2},\ldots,\theta_{n}) & = & \left(1+\prod_{i=1}^{n}S(\theta-\theta_{i})\right)F_{n}^{O}(\theta_{1},\theta_{2},\ldots,\theta_{n})\:,\label{eq:FFAxiom3SemiLocal}
\end{eqnarray}
and the general solution can be written as

\begin{equation}
\begin{array}{cc}
F_{n}(\theta_{1},\ldots,\theta_{n})= & H_{n}\left(\stackrel[i=1]{n}{\prod}\sqrt{x_{i}}\right)Q_{n}(x_{1},\ldots,x_{n})\stackrel[1\leqslant i<j\leqslant n]{}{\prod}\frac{f_{min}(\theta_{i}-\theta_{j})}{x_{i}+x_{j}}\end{array},\label{eq:FFParametrizationSemiLocal}
\end{equation}
where the parametrisation ensures that $Q_{n}$ is a symmetric polynomial
of the variables $\{x_{i}\}$ and that (\ref{eq:FFAxiom1SemiLocal})
and (\ref{eq:FFAxiom2SemiLocal}) are automatically satisfied as long
as choosing the Riemann sheets of the square root function according
to the prescriptions $\sqrt{e^{2\pi i}}=-1$ and $\sqrt{e^{\pi i}}=+i$
\footnote{It is also possible to switch to the other Riemann sheet by redefining
the form factor with appropriate signs.} From (\ref{eq:FFParametrizationSemiLocal}) and the Lorentz transformation
property (\ref{eq:RelInv}) it follows that for a semi-local operator
with integer Lorentz spin $s$ all matrix elements containing an odd
number of particles must vanish. Substituting (\ref{eq:FFParametrizationSemiLocal})
into (\ref{eq:FFAxiom3SemiLocal}) and fixing the normalisation as

\begin{equation}
H_{n+2}=H_{n}\mu^{2}\label{eq:SemiLocRecursionForNormConstant}
\end{equation}
with

\begin{equation}
\mu=\sqrt{\frac{4\sin\frac{\pi B}{2}}{f_{min}(i\pi)}}\:,\label{eq:Mu}
\end{equation}
a recursion equation can be written for $Q_{n}$

\begin{equation}
Q_{n+2}(-x,x,x_{1,}\ldots,x_{n})=C_{n}(x,x_{1},\ldots,x_{n})Q_{n}(x_{1,}\ldots,x_{n})\label{eq:C_n0A}
\end{equation}
with the kernel: 
\begin{equation}
\begin{array}{cc}
C_{n}(x,x_{1},\ldots,x_{n})= & \stackrel[k=0]{n}{\sum}\;\stackrel[m=0]{k}{\sum'}x^{2(n-k)+m}\sigma_{k}^{(n)}\sigma_{k-m}^{(n)}(-1)^{k+1}\;[m]_{c}\end{array}\:,\label{eq:C_n4}
\end{equation}
where $\sum'$ means summation for even indices only and we introduced
the notation

\begin{equation}
[m]_{c}=\begin{cases}
\frac{\cos\frac{\pi B}{2}m}{\sin\frac{\pi B}{2}} & \text{for }m\mathbb{\in Z}\backslash\{0\}\\
\frac{1}{2\sin\frac{\pi B}{2}} & \text{for }m=0
\end{cases}\:.\label{eq:M_C}
\end{equation}
Some details about the derivation of the above recursion are given
in Appendix \ref{sub:Appendix-B-Form-factor}.

\subsection{The sinh-Gordon twist field }

Similarly to the case of local operators, if a given $Q_{n}$ is a
solution of the recursion, then $Q'_{n}=\sigma_{n-1}^{(n)}\sigma_{1}^{(n)}Q_{n}$
is also a solution (corresponding to $\partial\bar{\partial}O$).
Therefore we focus on the so-called irreducible operators whose form
factors cannot be factorised. For irreducible semi-local operators
with Lorentz spin $s$ equal to zero, it follows from (\ref{eq:FFParametrizationSemiLocal})
that $Q_{0}$ and $Q_{2}$ must be a constant. In addition, a simple
power counting shows that the minimal solution $Q_{n}$ must have
a partial degree $n-2$ in each variable, since $C_{n}$ has a partial
degree $2$. The total degree can be determined from the Lorentz spin
zero condition with the result

\begin{eqnarray}
\text{Deg\ensuremath{_{tot}}}[Q_{n}] & = & \frac{n^{2}}{2}-n\label{eq:PolynomDegrees}\\
\text{Deg\ensuremath{_{part}}}[Q_{n}] & = & n-2\text{ for }n\geq2\:.\nonumber 
\end{eqnarray}
Setting $Q_{0}=1$, it follows immediately that $Q_{2}=-[0]_{c}$.
For $Q_{4}$, the most general symmetric polynomial of four variables
that satisfies irreducibility and (\ref{eq:PolynomDegrees}) is $A_{1}^{(4)}\sigma_{4}+A_{2}^{(4)}\sigma_{1}\sigma_{3}^{(4)}+A_{3}^{(4)}\sigma_{2}\sigma_{2}$\,.
Substituting this Ansatz into (\ref{eq:C_n4}), the unknown coefficients
$A_{k}^{(4)}$ can be determined. The solution for the coefficients
turns out to be unique: 
\begin{equation}
Q_{4}(x_{1},x_{2},x_{3},x_{4})=[0]_{c}\left([0]_{c}(\sigma_{1}\sigma_{3}+\sigma_{2}\sigma_{2})-([2]_{c}+2[0]_{c})\sigma_{4}\right)\:.\label{eq:Q_4Mod}
\end{equation}
$Q_{6}$ can be determined in a similar way with the result 
\begin{eqnarray}
Q_{6}(x_{1},..,x_{6}) & = & [0]_{c}\big(-[0]_{c}^{2}\sigma_{2}\sigma_{2}\sigma_{3}\sigma_{5}-[0]_{c}^{2}\sigma_{2}\sigma_{2}\sigma_{4}\sigma_{4}+[1]_{c}^{2}\sigma_{2}\sigma_{2}\sigma_{2}\sigma_{6}\label{eq:Q_6Mod}\\
 &  & -[0]_{c}^{2}\sigma_{1}\sigma_{3}\sigma_{4}\sigma_{4}+[1]_{c}^{2}\sigma_{4}\sigma_{4}\sigma_{4}-[0]_{c}^{2}\sigma_{1}\sigma_{3}\sigma_{3}\sigma_{5}+[0]_{c}^{2}\sigma_{3}\sigma_{4}\sigma_{5}\nonumber \\
 &  & +[0]_{c}([2]_{c}+[0]_{c})\sigma_{1}\sigma_{2}\sigma_{4}\sigma_{5}+[0]_{c}^{2}\sigma_{2}\sigma_{5}\sigma_{5}+[0]_{c}^{2}\sigma_{1}\sigma_{2}\sigma_{3}\sigma_{6}\nonumber \\
 &  & +[0]_{c}([2]_{c}+[0]_{c})\sigma_{3}\sigma_{3}\sigma_{6}-[0]_{c}([4]_{c}+5[2]_{c}+6[0]_{c})\sigma_{2}\sigma_{4}\sigma_{6}\nonumber \\
 &  & +[1]_{c}^{2}\sigma_{1}\sigma_{1}\sigma_{5}\sigma_{5}+[0]_{c}^{2}\sigma_{1}\sigma_{1}\sigma_{4}\sigma_{6}\nonumber \\
 &  & -[0]_{c}(2[4]_{c}+5[2]_{c}+7[0]_{c})\sigma_{1}\sigma_{5}\sigma_{6}\nonumber \\
 &  & +[0]_{c}([6]_{c}+3[4]_{c}+6[2]_{c}+7[0]_{c})\sigma_{6}\sigma_{6}\big)\:.\nonumber 
\end{eqnarray}
$Q_{8}$ has also been constructed and is again uniquely determined;
we omit its formula due to its length. In fact, the solution is unique
for any number of particles. The reason is that when determining $Q_{n}$
from $Q_{n-2}$, the ambiguity must be a symmetric polynomial $K_{n}$
satisfying
\begin{equation}
K_{n}(-x,x,x_{3},\dots,x_{n})=0
\end{equation}
However, any such polynomial can be written as 
\begin{equation}
K_{n}(x_{1},\dots,x_{n})=L_{n}(x_{1},\dots,x_{n})\prod_{i<j}(x_{i}+x_{j})
\end{equation}
where $L_{n}$ is an arbitrary symmetric polynomial. The degrees of
$K_{n}$ are 
\begin{align}
\text{Deg\ensuremath{_{tot}}}[K_{n}] & \geq\frac{n(n-1)}{2}\nonumber \\
\text{Deg\ensuremath{_{part}}}[K_{n}] & \geq n-1\text{ for }n\geq1\:,
\end{align}
which is higher than (\ref{eq:PolynomDegrees}), so no such ambiguity
exists for the minimal solution. The uniqueness of the solution is
in marked contrast to the case of local sinh-Gordon form factors where
the dimension of the linear space spanned grows by one for every level
of the recursion \cite{SinhGordonFF}. This can be understood considering
that the minimal local solutions are the exponential operators whose
space is spanned by the normal-ordered powers of the sinh-Gordon field
$\phi,$ so one indeed expects a countable infinity of independent
solutions. 

We denote the unique minimal spinless semi-local operator constructed
above by $\tau$. It can be checked explicitly that the form factors
of $\tau$ satisfy the cluster property (\ref{eq:ClusterProp}) and
therefore they are expected to correspond to a scaling field \cite{delta_theorem}.

The operator $\tau$ can be completely identified using the $\Delta$-theorem
sum rule (\ref{eq:DeltaTheoreM2}). Since the sinh-Gordon model has
a massive spectrum, the IR limit gives zero conformal weight. Therefore
(\ref{eq:DeltaTheoreM2}) equals the conformal weight of the operator
in the UV limiting theory, which is the massless free boson. Evaluating
the integral numerically for $n=2,4$ and $6$ and for various values
of the coupling strength $B$ the following results were obtained:

\begin{center}
\begin{tabular}{|c|c|c|c|}
\hline 
B  & 0.1  & 0.5  & 0.9\tabularnewline
\hline 
\hline 
$2$  & 0.0638324  & 0.067049  & 0.0681442\tabularnewline
\hline 
$2+4$  & 0.0624367  & 0.0618555  & 0.0613877\tabularnewline
\hline 
$2+4+6$  & 0.0624976  & 0.0627215  & 0.0625972\tabularnewline
\hline 
\end{tabular}
\par\end{center}

In the massless free bosonic theory, there exists an (up to normalisation)
unique field which is a conformal primary and changes the boundary
condition of the boson field from periodic to anti-periodic and vice
versa. This twist field has conformal weight $\Delta=1/16$ \cite{AppCFT}
and our considerations show that the above solution can be identified
with its off-critical version; the primary nature of the field corresponds
to the minimality of the solution; non-minimal semi-local solutions
of the bootstrap are expected to correspond to conformal descendants.

\subsection{Construction of the disorder operator}

We perform the roaming limit of the twist field the same way as discussed
in Section 2.3. The parametrisation for the form factors of $\tilde{\Phi}$
can be written as (\ref{eq:FFParametrizationFlow}). The square root
factors in the sinh-Gordon form factors of $\tau$ have the limits

\begin{equation}
\stackrel[i=1]{n}{\prod}\sqrt{x_{i}}\rightarrow e^{\theta_{0}r/2}\stackrel[i=1]{r}{\prod}\sqrt{x_{i}}\stackrel[j=1]{l}{\prod}\sqrt{y_{j}^{-1}}\:.\label{eq:SquareRootLimit}
\end{equation}
Due to the lack of a closed formula for the function $Q$ for the
twist field form factors, we cannot determine the dominant cells directly.
However, a direct generalisation of the cell structure for $\Theta$
and $\Phi$ works. We can use (\ref{eq:CellStep2}) for this case,
with the restriction that $l+r$ is even. Therefore, the dominant
cells to be considered are as follows: 
\begin{eqnarray*}
2\mbox{ particles} & : & [1,1]\quad[1,-1]\quad[-1,-1]\\
4\mbox{ particles} & : & [1,1,1,1]\quad[1,1,1,-1]\quad[1,1,-1,-1]\quad[1,-1,-1,-1]\quad[-1,-1,-1,-1]\\
\mbox{etc.}
\end{eqnarray*}
For $2$ particles, the scaling of the form factor constituents in
the roaming limit is as follows

\begin{center}
\begin{tabular}{|c|c|c|c|c|c|c|}
\hline 
$n=2$  & $\tilde{Q}_{r,l}$  & Normalisation  & $Q_{n}$  & Denominator  & Square root  & Overall scaling\tabularnewline
\hline 
\hline 
$r=0$  & $-1$  & $e^{\theta_{0}}$  & $e^{-\theta_{0}}$  & 1  & 1  & 1\tabularnewline
\hline 
$r=1$  & $-1$  & $e^{3\theta_{0}/2}$  & $e^{-\theta_{0}}$  & $e^{-\theta_{0}}$  & $e^{\theta_{0}/2}$  & 1\tabularnewline
\hline 
$r=2$  & $-1$  & $e^{\theta_{0}}$  & $e^{-\theta_{0}}$  & $e^{-\theta_{0}}$  & $e^{\theta_{0}}$  & 1\tabularnewline
\hline 
\end{tabular}
\par\end{center}

\noindent while for $4$ particles one has

\begin{center}
\begin{tabular}{|c|c|c|c|c|c|c|}
\hline 
$n=4$  & $\tilde{Q}_{r,l}$  & Normalisation  & $Q_{n}$  & Denominator  & Square root  & Overall scaling\tabularnewline
\hline 
\hline 
$r=0$  & $y_{1}^{2}y_{2}^{2}y_{3}^{2}y_{4}^{2}$  & 1  & 1  & 1  & 1  & 1\tabularnewline
\hline 
$r=1$  & $\frac{y_{1}y_{2}y_{3}}{x_{1}^{2}}$  & $e^{3\theta_{0}/2}$  & $e^{\theta}$  & $e^{-3\theta_{0}}$  & $e^{\theta_{0}/2}$  & 1\tabularnewline
\hline 
$r=2$  & $\frac{1}{x_{1}x_{2}}+y_{1}y_{2}$  & $e^{2\theta_{0}}$  & $e^{2\theta_{0}}$  & $e^{-5\theta_{0}}$  & $e^{\theta_{0}}$  & 1\tabularnewline
\hline 
$r=3$  & $\frac{1}{y_{1}}$  & $e^{3\theta_{0}/2}$  & $e^{3\theta_{0}}$  & $e^{-6\theta_{0}}$  & $e^{3\theta_{0}/2}$  & 1\tabularnewline
\hline 
$r=4$  & $x_{1}x_{2}x_{3}x_{4}$  & 1  & $e^{4\theta}$  & $e^{-6\theta_{0}}$  & $e^{2\theta_{0}}$  & 1\tabularnewline
\hline 
\end{tabular}
\par\end{center}

\noindent For the $6$ particle case the results are presented in
Appendix \ref{sub:Appendix-C-Tables}.

The exact form factors of $\tilde{\Phi}$ resulting from the massless
form factor bootstrap are given by (\cite{MasslessFlow}, cf.\ also
Appendix \ref{sub:Appendix-A-MasslessFF}) 
\begin{equation}
\begin{array}{ll}
Q_{r,0}=\rho_{r}^{(r-1)/2} & \text{for even }r\\
Q_{r,1}=\frac{\rho_{r}^{r/2-1}}{\lambda_{1}^{r/2}} & \text{for odd }r\\
Q_{r,2}=\rho_{r}^{(r-3)/2}\stackrel[k=0]{r}{\sum'}\rho_{k}\lambda_{2}^{(k-r+1)/2} & \text{for even }r\\
Q_{r,3}=\frac{\rho_{r}^{r/2-2}}{\lambda_{3}^{r/2-1}}\stackrel[k=0]{r}{\sum'}\rho_{k}\lambda_{2}^{k/2} & \text{for odd }r,
\end{array}\label{eq:QDisorderRight}
\end{equation}
in which the primed sum means summation on even indices and which
are indeed equal to the corresponding $\tilde{Q}$ functions multiplied
by the appropriate square root factors.

A more careful examination shows that for $n=2$ and $6$, there is
a sign difference between the limit of the twist field $Q$ functions
and the exact expression (\ref{eq:QDisorderRight}). This alternating
sign is, however, cancelled by the normalisation factors. The normalization
of the massless bootstrap solution reads

\begin{eqnarray}
H_{r,l} & = & 2^{r(r-1)/2+l(l-1)/2}\gamma^{-rl/2}i^{(rl+r+l)/2}\quad\text{for \ensuremath{r,l} even }\label{eq:QDisorderRightNomr}\\
H_{r,l} & = & \sqrt{2}2^{r(r-1)/2+l(l-1)/2}\gamma^{-(rl+1)/2}i^{(rl+r+l-3)/2}\quad\text{for \ensuremath{r,l} odd ,}\nonumber 
\end{eqnarray}
which corresponds to $F_{0,0}=1$ and $F_{1,1}=1$ in the IR limit
$\theta\rightarrow0$. It is important to stress that the recursion
equation separates into two subsystems, one of which relates form
factors of the form $F_{2k,2l}$, while the other subsystem couples
form factors of the form $F_{2k+1,2l+1}$. Due to this decoupling,
it is possible and even necessary to apply a slightly different normalisation
for the cases $r,l$ even/odd when performing the roaming limit of
$\tau$. The limit of the prefactors for the form factors of the twist
field are

\begin{equation}
\begin{array}{cc}
2^{r(r-1)/2+l(l-1)/2}\gamma^{-rl/2}i^{(rl+r+l)/2}i^{-(n^{2})/2} & \text{for \ensuremath{r,l} even }\\
2^{r(r-1)/2+l(l-1)/2}\gamma^{-(rl+1)/2}i^{(rl+r+l-3)/2}i^{-(n^{2}-3)/2} & \text{for \ensuremath{r,l} odd .}
\end{array}\label{eq:TwistPreftNomr}
\end{equation}
For even $n$, $(-i)^{\frac{n^{2}}{2}}$ is $+1$ if $n=0,4,8,...$,
and $-1$ if $n=2,6,10,...$. This alternating sign cancels out the
$\pm1$ factors from the limit of the polynomial of the twist field,
hence the properly normalised limit of the sinh-Gordon form factors
is defined as

\begin{equation}
F_{r,l}^{\tilde{\Phi}}=\underset{\theta_{0}\rightarrow\infty}{\lim}F_{n}^{\tau}(\theta_{1}+\theta_{0},\ldots,\theta_{r}+\theta_{0},\theta'_{1},\ldots,\theta'_{l},B(\pm\theta_{0}))\ \text{for \ensuremath{r,l} even }\:,\label{eq:TwistFieldLimitEvenEven}
\end{equation}
and

\begin{equation}
F_{r,l}^{\tilde{\Phi}}=i^{-3/2}\sqrt{2}\gamma^{-1/2}\underset{\theta_{0}\rightarrow\infty}{\lim}F_{n}^{\tau}(\theta_{1}+\theta_{0},\ldots,\theta_{r}+\theta_{0},\theta'_{1},\ldots,\theta'_{l},B(\pm\theta_{0}))\:\text{for \ensuremath{r,l} odd \:.}\label{eq:TwistFieldLimitOddOdd}
\end{equation}
Note that the roaming limit of the twist field is independent on the
sign of $\theta_{0}$, therefore the CPT symmetrisation is trivial
for the case of the disorder operator. 

As a further test of the roaming limit construction, it is instructive
to check if $\tau$ reproduces the  disorder operator in step $k=1$,
i.e., the perturbed Ising model. Indeed, with the first few solutions
of $Q^{\tau}$ it is easy to check explicitly that

\begin{equation}
\underset{\theta_{0}\rightarrow\infty}{\lim}F_{n}^{\tau}(\theta_{1},\ldots,\theta_{n},B(\pm\theta_{0}))=i^{n/2}\stackrel[1\leq i<j\leq n]{}{\prod}\tanh(\frac{\theta_{i}-\theta{}_{j}}{2})\:,
\end{equation}
which are the form factors of the disorder operator in the perturbed
Ising model with vacuum expectation value normalized to unity \cite{YurovZamIsing}.

\section{Conclusions \label{sec:Conclusions}}

In this paper we have shown that it is possible to obtain the form
factors of the order and disorder operators of the massless RG flow
interpolating between the tricritical and the critical Ising model
as a properly defined roaming limit of certain sinh-Gordon form factors.
These results are a natural next step after the construction of the
form factors of the trace of the stress-energy tensor in \cite{RefRoaming},
which was based on Zamolodchikov's original staircase idea \cite{RefZamo}.
Our results demonstrate that the roaming construction of form factors
can be extended to other operators along the massless flows generated
by the roaming limit. We explicitly matched the resulting form factors
with the previously known explicit solutions \cite{MasslessFlow}.
For the order operator it is important to note that the roaming limit
explicitly constructs all the form factors for any number of particles,
while previously it was only possible to obtain form factors recursively.
In this construction we also found that in general, the roaming limit
must incorporate a CPT symmetrisation. From our field theoretic arguments
we expect that this lesson is general for any future extensions of
the roaming form factor approach.

The construction of the disorder field necessitated the consideration
of semi-local operators in the sinh-Gordon model. We found that the
minimal solution was unique, which is consistent with the identification
of the operator with the twist field changing the boundary conditions
of the boson from periodic to anti-periodic and vice versa. This identification
is strongly supported by the estimate of the scaling dimension obtained
from the $\Delta$-theorem.

There are several interesting open directions. Focusing on the massless
flow, it is possible to construct the roaming limit for other operators.
In principle, by solving the form factor bootstrap for the minimal
model flow along the lines of \cite{MasslessFlow} at least the first
few form factors of these operators can be constructed for the flow
enabling a direct comparison with the limit of sinh-Gordon form factors,
which can support the validity of the limiting procedure. For operators
whose form factors are non-vanishing for even particle numbers, the
$\Delta$-theorem can help to identify the operator. An even more
interesting issue is to extend the construction of form factors of
various operators for flows between higher minimal models corresponding
to higher steps in Zamolodchikov's staircase. As a consequence of
the non-diagonal scattering, for higher steps the solution of the
FF bootstrap equation is highly non-trivial and is presently unknown.
Therefore the construction of form factors with the roaming approach
seems very promising, even if the magnonic structure of the limit
\cite{RefRoaming} indicates a non-trivial relation to form factors
being written in the standard RSOS kink basis \cite{ReshetikinSmirnov,masslessRSOS}.
Therefore the main issue to be solved is the proper treatment of the
magnonic structure in the roaming limit. An even more ambitious goal
is to extend the construction to other known staircase flows, and
to connect the form factors of their local operators with the Dynkin
formulation of the corresponding magnonic TBA systems \cite{Ravanini:1992fi},
as suggested in \cite{RefRoaming}.

\subsection*{Acknowledgements}

This research was supported by the Momentum grant LP2012-50 of the
Hungarian Academy of Sciences, by the STFC under grant number ST/L000407/1,
and by the Marie Curie network GATIS (gatis.desy.eu) of the European
Union's Seventh Framework Programme FP7/2007-2013/ under REA Grant
Agreement No 317089. We are grateful to István Szécsényi for helpful
discussions and collaboration on related matters. 

\appendix

\section{Form factor bootstrap for the massless flow\label{sub:Appendix-A-MasslessFF}}

In the following we briefly review the main steps of the form factor
bootstrap for the QFT associated with the RG flow between the two
fixed points. These results were previously obtained in \cite{MasslessFlow};
however, matching the form factors to the roaming limit requires changes
in the normalisation conventions, so we briefly give the necessary
details here.

\subsection{The recursion relations for the form factor polynomials}

As discussed in the main text, the spectrum of the theory consists
of right-moving and left-moving particles with dispersion $p^{0}=p^{1}=\frac{M}{2}e^{\theta}$/$p^{0}=-p^{1}=\frac{M}{2}e^{-\theta}$,
whereas with the massless creation and annihilation operators the
asymptotic in- and out-states can be written similarly to the massive
cases. The massless S matrices \cite{MasslessFlow} are

\begin{equation}
\begin{array}{c}
S_{LL}(\theta)=S_{RR}(\theta)=-1\\
S_{RL}(\theta)=\tanh(\theta/2-i\pi/4)\:.
\end{array}\label{eq:SMatrixFlow}
\end{equation}
The form factors of the theory associated with the RG flow are defined
as 
\[
F_{\alpha_{1},\ldots,\alpha_{n}}^{O}(\theta_{1},\ldots,\theta_{n})=\langle0|O(0,0)|A_{\alpha_{1}}(\theta_{1})\ldots A_{\alpha_{n}}(\theta_{n})\rangle
\]
and must satisfy the equations

\begin{equation}
F_{\alpha_{1},\ldots,\alpha_{i},\alpha_{i+1},\ldots\alpha_{n}}^{O}(\theta_{1},\ldots\theta_{i},\theta_{i+1},\ldots,\theta_{n})=S_{\alpha_{i},\alpha_{i+1}}(\theta_{i}-\theta_{i+1})F_{\alpha_{1},\ldots,\alpha_{i+1},\alpha_{i},\ldots\alpha_{n}}^{O}(\theta_{1},\ldots\theta_{i+1},\theta_{i},\ldots,\theta_{n})\label{eq:FFAxiom1App}
\end{equation}

\begin{equation}
F_{\alpha_{1},\alpha_{2},\ldots\alpha_{n}}^{O}(\theta_{1}+2\pi i,\theta_{2},\ldots,\theta_{n})=\pm F_{\alpha_{2},\ldots\alpha_{n},\alpha_{1}}^{O}(\theta_{2},\ldots,\theta_{n},\theta_{1})\label{eq:FFAxiom2App}
\end{equation}

\begin{equation}
-i\underset{\theta'=\theta+i\pi}{\mbox{Res}}F_{\alpha,\alpha,\alpha_{1},\ldots\alpha_{n}}^{O}(\theta',\theta,\theta_{1},\theta_{2},\ldots,\theta_{n})=\left(1\mp\prod_{i=1}^{n}S_{\alpha,\alpha_{i}}(\theta-\theta_{i})\right)F_{\alpha_{1},\ldots\alpha_{n}}^{O}(\theta_{1},\theta_{2},\ldots,\theta_{n})\:,\label{eq:FFAxiom3App}
\end{equation}
where $\alpha$ and $\alpha_{i}$ denotes either a right-moving (R)
or a left-moving (L) particle. The choice of signs in the cyclic permutation
and kinematic pole equations corresponds to local (upper choice) or
semi-local (lower choice) operators, the latter being relevant for
the disorder field. The form factors are parametrised as

\begin{equation}
\begin{array}{cc}
F_{r,l}(\theta_{1},\ldots,\theta_{r},\theta'_{1},\ldots,\theta'_{l})= & H_{r,l}Q_{r,l}(x_{1},\ldots,x_{r},y_{1}\ldots y_{l})\underset{1\leqslant i<j\leqslant r}{\prod}\frac{f_{RR}(\theta_{i}-\theta_{j})}{x_{i}+x_{j}}\\
 & \times\stackrel[i=1]{r}{\prod}\stackrel[j=1]{l}{\prod}f_{RL}(\theta_{i}-\theta'_{j})\underset{1\leqslant i<j\leqslant l}{\prod}\frac{f_{LL}(\theta'_{i}-\theta'_{j})}{y_{i}+y_{j}}\:,
\end{array}\label{eq:FFParametrizationFlowApp}
\end{equation}
where $r,l$ are the number of right/left movers, $H_{r,l}$ is a
normalisation constant, $Q_{r,l}$ are meromorphic function of variables
$x_{i}=e^{\theta_{i}}$ and $y_{i}=e^{-\theta'_{i}}$ and the functions
$f_{RR}$, $f_{LL}$ and $f_{RL}$ are the minimal form factors $f_{RR}(\theta)=f_{LL}(\theta)=\sinh(\theta/2)$
and 
\begin{equation}
f_{RL}(\theta)=\exp\left(\frac{\theta}{4}-\int_{0}^{\infty}\frac{dt}{t}\frac{\sin^{2}(\frac{i\pi-\theta}{2\pi}t)}{\sinh(t)\cosh(t/2)}\right)\:,\label{eq:fRLApp}
\end{equation}
which solve 
\begin{equation}
f_{\alpha,\beta}(\theta)=S_{\alpha,\beta}(\theta)f_{\alpha,\beta}(\theta+2\pi i)\:.\label{eq:MonodromyPropertiesApp}
\end{equation}
In addition, function $f_{RL}(\theta)$ satisfies

\begin{equation}
f_{RL}(\theta\pm i\pi)f_{RL}(\theta)=i\gamma(1\pm ie^{-\theta})^{-1}\:,\label{eq:fRLPi}
\end{equation}
where $\gamma=\sqrt{2}e^{2K/\pi}$ and $K$ is Catalan's constant.
For the disorder operator the $Q_{r,l}$ functions contain a factor
\[
\sqrt{\stackrel[i=1]{r}{\prod}x_{i}\stackrel[j=1]{l}{\prod}y_{j}}\:.
\]
The kinematical pole equation (\ref{eq:FFAxiom3App}) gives a recurrence
relation for the polynomials. With the normalisation

\begin{equation}
H_{r+2,l}=i^{l+1}2^{2r+1}\gamma^{-l}H_{r,l}\:,\label{eq:ReqRNorm}
\end{equation}
the right-mover recurrence relations for the polynomials read

\begin{equation}
Q_{r+2,l}(-x,x,x_{1,}\ldots,y_{1},\ldots)=i^{r-l+1}x^{r-l+1}\frac{\rho_{r}}{\lambda_{l}}\stackrel[k=0]{l}{\sum'}(-ix)^{k}\lambda_{k}(\{y_{j}\})Q_{r,l}\:,\label{eq:ReqR}
\end{equation}
where $\rho_{r}$ denotes the $r$th elementary symmetric polynomial
of variables $x_{1},\ldots,x_{r}$ and $\lambda_{l}$ denotes the
$l$th elementary symmetric polynomial of variables $y_{1},\ldots,y_{l}$
. For the left-movers one obtains 
\begin{equation}
Q_{r,l+2}(,x\ldots,y_{1},\ldots,y_{l},y,-y)=(-i)^{l-r+1}y^{l-r+1}\frac{\lambda_{l}}{\rho_{r}}\stackrel[k=0]{r}{\sum'}(-iy)^{k}\rho_{k}(\{x_{i}\})Q_{r,l}\label{eq:ReqL}
\end{equation}
with the choice 
\begin{equation}
H_{r,l+2}=i^{r+1}2^{2l+1}\gamma^{-r}H_{r,l}\:.\label{eq:ReqLNorm}
\end{equation}

Equations (\ref{eq:ReqR}) and (\ref{eq:ReqL}) differ from the corresponding
equations in \cite{MasslessFlow} by some powers of $i$. As a consequence,
the recursion relation for the normalisation constants is also slightly
modified.

\subsection{Solution for the order operator}

For the order operator we require the normalisation $F_{1,0}=F_{0.1}=1$
, or equivalently $Q_{1,0}=Q_{0.1}=1$ . Then the solution for the
$Q$ is \cite{MasslessFlow}

\begin{equation}
\begin{array}{ll}
Q_{r,0}=\rho_{r}^{(r-1)/2} & \text{for odd }r\\
Q_{r,1}=\frac{\rho_{r}^{r/2-1}}{\lambda_{1}^{r/2}} & \text{for even }r\\
Q_{r,2}=\rho_{r}^{(r-3)/2}\stackrel[k=0]{r}{\sum'}\rho_{k}\lambda_{2}^{(k-r+1)/2} & \text{for odd }r\\
Q_{r,3}=\frac{\rho_{r}^{r/2-2}}{\lambda_{3}^{r/2-1}}\stackrel[k=0]{r}{\sum'}\rho_{k}\lambda_{2}^{k/2} & \text{for even }r\:,
\end{array}\label{eq:QOrderRightApp}
\end{equation}
and the primed sum means summation on even indices. To obtain the
functions $Q_{0,l}$, $Q_{1,l}$, $Q_{2,l}$ and $Q_{3,l}$ one simply
needs to exchange the elementary symmetric polynomials $\rho\longleftrightarrow\lambda$
in (\ref{eq:QOrderRightApp}). The explicit normalisation constants
are

\begin{equation}
H_{r,l}=2^{r(r-1)/2+l(l-1)/2}\gamma^{-rl/2}i^{(rl+r+l-1)/2}\:.\label{eq:QOrderRightNomr}
\end{equation}

\subsection{Solution for the disorder operator}

For the disorder operator $\widetilde{\Phi}$, we stipulate $F_{0.0}=1$
and also $F_{1,1}=1$ in the IR limit, i.e. when the rapidity difference
in the form factor tends to $0$. From these constraints we have $Q_{0,0}=1$
and 
\[
Q_{1,1}=\frac{1}{\rho_{1}^{1/2}\lambda_{1}^{1/2}}\,,
\]
where the choice of $Q_{1,1}$ is made unique by requiring that the
operator has a zero Lorentz spin and is symmetric under parity that
swaps left-movers with right-movers. With these initial conditions,
the $Q_{r,l}$ functions are the following \cite{MasslessFlow}

\begin{equation}
\begin{array}{ll}
Q_{r,0}=\rho_{r}^{(r-1)/2} & \text{for even }r\\
Q_{r,1}=\frac{\rho_{r}^{r/2-1}}{\lambda_{1}^{r/2}} & \text{for odd }r\\
Q_{r,2}=\rho_{r}^{(r-3)/2}\stackrel[k=0]{r}{\sum'}\rho_{k}\lambda_{2}^{(k-r+1)/2} & \text{for even }r\\
Q_{r,3}=\frac{\rho_{r}^{r/2-2}}{\lambda_{3}^{r/2-1}}\stackrel[k=0]{r}{\sum'}\rho_{k}\lambda_{2}^{k/2} & \text{for odd }r,
\end{array}\label{eq:QDisorderRightApp}
\end{equation}
in which the primed sum means summation on even indices and which
contain the square root type products that are necessary to fulfil
(\ref{eq:FFAxiom2App}). These functions, again, are consistent with
(\ref{eq:ReqR}). Similarly to (\ref{eq:QOrderRightApp}), for the
functions $Q_{0,l}$, $Q_{1,l}$, $Q_{2,l}$ and $Q_{3,l}$ one simply
needs to exchange the elementary symmetric polynomials $\rho\longleftrightarrow\lambda$
in (\ref{eq:QDisorderRightApp}).

The normalisation is chosen in a slightly different way for odd-odd
and even-even form factors:

\begin{equation}
\begin{array}{cc}
H_{r,l}=2^{r(r-1)/2+l(l-1)/2}\gamma^{-rl/2}i^{(rl+r+l)/2} & \text{for both \ensuremath{r,l} even }\\
H_{r,l}=\sqrt{2}2^{r(r-1)/2+l(l-1)/2}\gamma^{-(rl+1)/2}i^{(rl+r+l-3)/2} & \text{for both \ensuremath{r,l} odd \:.}
\end{array}\label{eq:QDisorderRightNomrApp}
\end{equation}

\section{Form factor bootstrap for semi-local fields\label{sub:Appendix-B-Form-factor}}

In view of the lack of bound state particles in the sinh-Gordon model,
the starting equations for the bootstrap are

\begin{equation}
F_{n}^{O}(\theta_{1},\ldots\theta_{i},\theta_{i+1},\ldots,\theta_{n})=S(\theta_{i}-\theta_{i+1})F_{n}^{O}(\theta_{1},\ldots\theta_{i+1},\theta_{i},\ldots,\theta_{n})\label{eq:FFAxiom1SemiLocal-1}
\end{equation}

\begin{equation}
F_{n}^{O}(\theta_{1}+2\pi i,\theta_{2},\ldots,\theta_{n})=-F_{n}^{O}(\theta_{2},\ldots,\theta_{n},\theta_{1})\label{eq:FFAxiom2SemiLocal-1}
\end{equation}

\begin{equation}
-i\underset{\theta'=\theta+i\pi}{\mbox{Res}}F_{n}^{O}(\theta',\theta,\theta_{1},\theta_{2},\ldots,\theta_{n})=\left(1+\prod_{i=1}^{n}S(\theta-\theta_{i})\right)F_{n}^{O}(\theta_{1},\theta_{2},\ldots,\theta_{n})\:.\label{eq:FFAxiom3SemiLocal-1}
\end{equation}
Hence the most general form of such form factors can be written

\begin{equation}
\begin{array}{cc}
F_{n}(\theta_{1},\ldots,\theta_{n})= & H_{n}\left(\stackrel[i=1]{n}{\prod}\sqrt{x_{i}}\right)Q_{n}(x_{1},\ldots,x_{n})\stackrel[1\leqslant i<j\leqslant n]{}{\prod}\frac{f_{min}(\theta_{i}-\theta_{j})}{x_{i}+x_{j}}\end{array},\label{eq:FFParametrizationSemiLocal-1}
\end{equation}
where $Q_{n}$ is a symmetric polynomial of the variables $\{x_{i}\}$;
this satisfies equations (\ref{eq:FFAxiom1SemiLocal-1})-(\ref{eq:FFAxiom2SemiLocal-1}).

Plugging (\ref{eq:FFParametrizationSemiLocal-1}) into the kinematical
singularity equation (\ref{eq:FFAxiom3SemiLocal-1}) gives a recurrence
relation

\begin{eqnarray}
 &  & -H_{n+2}Q_{n+2}(-x,x,x_{1,}\ldots,x_{n})f_{min}(i\pi)\stackrel[i=1]{n}{\prod}\frac{f_{min}(\theta+i\pi-\theta_{i})f_{min}(\theta-\theta_{i})}{x_{i}^{2}-x^{2}}\label{eq:RecStartSL1}\\
 &  & =H_{n}\left(1+\stackrel[i=1]{n}{\prod}S(\theta-\theta_{i})\right)Q_{n}(x_{1},\ldots,x_{n}).\nonumber 
\end{eqnarray}
Using the identity

\begin{equation}
f_{min}(\theta+i\pi)f_{min}(\theta)=\frac{\sinh(\theta)}{\sinh(\theta)+(\omega-\omega^{-1})/2}\:,\label{eq:SinhGordonfmin}
\end{equation}
and also

\[
\left(1+\stackrel[i=1]{n}{\prod}S(\theta-\theta_{i})\right)=\frac{\stackrel[i=1]{n}{\prod}x^{2}-x_{i}^{2}+xx_{i}(\omega-\omega^{-1})+\stackrel[i=1]{n}{\prod}x^{2}-x_{i}^{2}-xx_{i}(\omega-\omega^{-1})}{\stackrel[i=1]{n}{\prod}x^{2}-x_{i}^{2}+xx_{i}(\omega-\omega^{-1})}\:,
\]
resulting from (\ref{eq:SinhGordonS}), the recursion can be written
as 
\begin{eqnarray}
 &  & -(-1)^{n}H_{n+2}f_{min}(i\pi)Q_{n+2}(-x,x,x_{1,}\ldots,x_{n})=\label{eq:RecStartSL2}\\
 &  & \left[\stackrel[i=1]{n}{\prod}x^{2}-x_{i}^{2}+xx_{i}(\omega-\omega^{-1})+\stackrel[i=1]{n}{\prod}x^{2}-x_{i}^{2}-xx_{i}(\omega-\omega^{-1})\right]H_{n}Q_{n}\:.\nonumber 
\end{eqnarray}
where $(-1)^{n}=1$ since $n$ is even. Introducing the quantity 
\[
\mu=\sqrt{\frac{4\sin\frac{\pi B}{2}}{f_{min}(i\pi)}}\:,
\]
we can define 
\[
H_{n+2}=H_{n}\mu^{2}\:.
\]
Then the recurrence relation takes the form 
\begin{equation}
Q_{n+2}(-x,x,x_{1,}\ldots,x_{n})=\frac{-1}{4\sin\frac{\pi B}{2}}\left[\stackrel[i=1]{n}{\prod}x^{2}-x_{i}^{2}+xx_{i}(\omega-\omega^{-1})+\stackrel[i=1]{n}{\prod}x^{2}-x_{i}^{2}-xx_{i}(\omega-\omega^{-1})\right]Q_{n}\:,\label{eq:C_n00}
\end{equation}
which can be simplified to 
\begin{equation}
Q_{n+2}(-x,x,x_{1,}\ldots,x_{n})=C_{n}(x,x_{1},\ldots,x_{n})Q_{n}(x_{1,}\ldots,x_{n})\:,\label{eq:C_n0B}
\end{equation}
where 
\begin{equation}
C_{n}(x,x_{1},\ldots,x_{n})=\frac{-1}{4\sin\frac{\pi B}{2}}\left[\stackrel[i=1]{n}{\prod}(x+\omega x_{i})(x-\omega^{-1}x_{i})+\stackrel[i=1]{n}{\prod}(x-\omega x_{i})(x+\omega^{-1}x_{i})\right]\:.\label{eq:C_n1}
\end{equation}
We proceed by manipulating $C_{n}$. Using the generator function
of the elementary symmetric polynomials $\stackrel[i=1]{n}{\prod}(x+\omega x_{i})=\stackrel[k=0]{n}{\sum}\omega^{k}x^{n-k}\sigma_{k}^{(n)}$
and $\stackrel[i=1]{n}{\prod}(x-\omega^{-1}x_{i})=\stackrel[l=0]{n}{\sum}\omega^{-l}(-1)^{l}x^{n-l}\sigma_{l}^{(n)}$
and exploiting the fact that the two products in (\ref{eq:C_n1})
differ only by $\omega\rightarrow-\omega$ yields 
\begin{equation}
C_{n}(x,x_{1},\ldots,x_{n})=\frac{-1}{4\sin\frac{\pi B}{2}}\left[\stackrel[k=0]{n}{\sum}\;\stackrel[l=0]{n}{\sum}\omega^{k-l}x^{2n-k-l}\sigma_{k}^{(n)}\sigma_{l}^{(n)}(1+(-1)^{k+l})(-1)^{l}\right]\:.\label{eq:C_n2}
\end{equation}
Introducing the functions $[m]_{c}$

\begin{equation}
[m]_{c}=\begin{cases}
\frac{\cos\frac{\pi B}{2}m}{\sin\frac{\pi B}{2}} & \text{for }m\mathbb{\in Z}\backslash\{0\}\\
\frac{1}{2\sin\frac{\pi B}{2}} & \text{for }m=0
\end{cases}\:,\label{eq:M_C-1}
\end{equation}
and rearranging the sums our final form for the recursion kernel is

\begin{equation}
\begin{array}{cc}
C_{n}(x,x_{1},\ldots,x_{n})= & \stackrel[k=0]{n}{\sum}\;\stackrel[m=0]{k}{\sum'}x^{2(n-k)+m}\sigma_{k}^{(n)}\sigma_{k-m}^{(n)}(-1)^{k+1}\;[m]_{c}\end{array}\:.\label{eq:Cn_4B}
\end{equation}

\section{Limits of $5$ and $6$ particle form factor polynomials \label{sub:Appendix-C-Tables}}

In the following, the roaming limits $\tilde{Q}_{r,l}$ resulting
from the polynomials $Q_{5}^{1/2}$ are presented. Normalising them
with $F_{1}=\frac{1\pm i}{\sqrt{2}}e^{\theta_{0}/4}$ and performing
the CPT symmetrisation that eliminates the imaginary terms, they are
found to be identical with (\ref{eq:QOrderRight}):

\begin{equation}
\begin{array}{ll}
{\tilde{Q}_{0,5}} & {=(y_{1}y_{2}y_{3}y_{4}y_{5})^{2}}\\
{} & {}\\
{\tilde{Q}_{1,4}} & {=y_{1}y_{2}y_{3}y_{4}\left(\frac{1}{x_{1}^{2}}\pm\frac{i(y_{1}+y_{2}+y_{3}+y_{4})}{x_{1}}\right)}\\
{} & {}\\
{\tilde{Q}_{2,3}} & {=\frac{1+x_{1}x_{2}(y_{2}y_{3}+y_{1}y_{3}+y_{1}y_{2})\pm i\left(x_{2}x_{1}^{2}y_{1}y_{2}y_{3}+x_{1}\left(y_{1}\left(x_{2}^{2}y_{2}y_{3}+1\right)+y_{2}+y_{3}\right)+x_{2}\left(y_{1}+y_{2}+y_{3}\right)\right)}{x_{1}x_{2}}}\\
{} & {}\\
{\tilde{Q}_{3,2}} & {=\frac{1+x_{1}x_{2}y_{1}y_{2}+x_{1}x_{3}y_{1}y_{2}+x_{2}x_{3}y_{1}y_{2}\pm i\left(y_{1}+y_{2}\right)\left(x_{1}\left(x_{2}x_{3}y_{1}y_{2}+1\right)+x_{2}+x_{3}\right)}{y_{1}y_{2}}}\\
{} & {}\\
{\tilde{Q}_{4,1}} & {=\frac{x_{1}x_{2}x_{3}x_{4}\left(1\pm i(x_{1}y_{1}+x_{2}y_{1}+x_{3}y_{1}+x_{4}y_{1})\right)}{y_{1}^{2}}}\\
{} & {}\\
{\tilde{Q}_{5,0}} & {=x_{1}^{2}x_{2}^{2}x_{3}^{2}x_{4}^{2}x_{5}^{2}\:.}\\
\\
\end{array}
\end{equation}
From the following table, the asymptotic divergence of various parts
of the sinh-Gordon form factors can be studied. The overall scaling
is $e^{\theta_{0}/4}$ again as predicted by (\ref{eq:Nodd}).

\begin{center}
\begin{tabular}{|c|c|c|c|c|}
\hline 
$n=5$  & Normalisation  & $Q_{5}^{1/2}$  & Denominator  & Overall scaling\tabularnewline
\hline 
\hline 
$r=0$  & $e^{-5\theta_{0}/4}$  & $e^{2\theta_{0}}$  & 1  & $e^{3\theta_{0}/4}$\tabularnewline
\hline 
$r=1$  & $e^{3\theta_{0}/4}$  & $e^{4\theta_{0}}$  & $e^{-4\theta_{0}}$  & $e^{3\theta_{0}/4}$\tabularnewline
\hline 
$r=2$  & $e^{7\theta_{0}/4}$  & $e^{6\theta_{0}}$  & $e^{-7\theta_{0}}$  & $e^{3\theta_{0}/4}$\tabularnewline
\hline 
$r=3$  & $e^{7\theta_{0}/4}$  & $e^{8\theta_{0}}$  & $e^{-9\theta_{0}}$  & $e^{3\theta_{0}/4}$\tabularnewline
\hline 
$r=4$  & $e^{3\theta_{0}/4}$  & $e^{10\theta_{0}}$  & $e^{-10\theta_{0}}$  & $e^{3\theta_{0}/4}$\tabularnewline
\hline 
$r=5$  & $e^{-5\theta_{0}/4}$  & $e^{12\theta_{0}}$  & $e^{-10\theta_{0}}$  & $e^{3\theta_{0}/4}$\tabularnewline
\hline 
\end{tabular}
\par\end{center}

Considering the limit of the polynomial $Q_{6}^{\tau}$ and multiplying
them with the square root type factors we obtain the functions (\ref{eq:QDisorderRight})
apart from a sign which is cancelled by another minus sign from the
normalisation.

\begin{equation}
\begin{array}{ccl}
\tilde{Q}_{0,6} & = & -y_{1}^{3}y_{2}^{3}y_{3}^{3}y_{4}^{3}y_{5}^{3}y_{6}^{3}\\
\\
\tilde{Q}_{1,5} & = & -\frac{y_{1}^{2}y_{2}^{2}y_{3}^{2}y_{4}^{2}y_{5}^{2}}{x_{1}^{3}}\\
\\
\tilde{Q}_{2,4} & = & -\frac{y_{1}y_{2}y_{3}y_{4}\left(x_{1}^{2}x_{2}^{2}y_{1}y_{2}y_{3}y_{4}+x_{1}x_{2}\left(y_{3}y_{4}+y_{2}\left(y_{3}+y_{4}\right)+y_{1}\left(y_{2}+y_{3}+y_{4}\right)\right)+1\right)}{x_{1}^{2}x_{2}^{2}}\\
\\
\tilde{Q}_{3,4} & = & -\frac{x_{2}x_{3}\left(y_{2}y_{3}+y_{1}\left(y_{2}+y_{3}\right)\right)+x_{1}\left(x_{2}+x_{3}\right)\left(y_{2}y_{3}+y_{1}\left(y_{2}+y_{3}\right)\right)+1}{x_{1}x_{2}x_{3}}\\
\\
\tilde{Q}_{4,2} & = & -\frac{x_{3}x_{4}y_{1}y_{2}+x_{2}\left(x_{3}+x_{4}\right)y_{1}y_{2}+x_{1}y_{1}y_{2}\left(x_{2}\left(x_{3}x_{4}y_{1}y_{2}+1\right)+x_{3}+x_{4}\right)+1}{y_{1}y_{2}}\\
\\
\tilde{Q}_{5,1} & = & -\frac{x_{1}x_{2}x_{3}x_{4}x_{5}}{y_{1}^{2}}\\
\\
\tilde{Q}_{6,0} & = & -x_{1}^{2}x_{2}^{2}x_{3}^{2}x_{4}^{2}x_{5}^{2}x_{6}^{2}\:.
\end{array}
\end{equation}
The overall scaling of the form factors is identical with scaling
of $F_{4}^{\tau}$ as shown in the following table:

\begin{center}
\begin{tabular}{|c|c|c|c|c|c|}
\hline 
$n=6$  & Normalisation  & $Q_{n}$  & Denominator  & Square root  & Overall scaling\tabularnewline
\hline 
\hline 
$r=0$  & $e^{-3\theta_{0}}$  & $e^{3\theta_{0}}$  & 1  & 1  & 1\tabularnewline
\hline 
$r=1$  & $e^{-\theta_{0}/2}$  & $e^{5\theta}$  & $e^{-5\theta_{0}}$  & $e^{\theta_{0}/2}$  & 1\tabularnewline
\hline 
$r=2$  & $e^{\theta_{0}}$  & $e^{7\theta_{0}}$  & $e^{-9\theta_{0}}$  & $e^{\theta_{0}}$  & 1\tabularnewline
\hline 
$r=3$  & $e^{3\theta_{0}/2}$  & $e^{9\theta_{0}}$  & $e^{-12\theta_{0}}$  & $e^{3\theta_{0}/2}$  & 1\tabularnewline
\hline 
$r=4$  & $e^{\theta_{0}}$  & $e^{11\theta_{0}}$  & $e^{-14\theta_{0}}$  & $e^{2\theta_{0}}$  & 1\tabularnewline
\hline 
$r=5$  & $e^{-\theta_{0}/2}$  & $e^{13\theta_{0}}$  & $e^{-15\theta_{0}}$  & $e^{5\theta_{0}/2}$  & 1\tabularnewline
\hline 
$r=6$  & $e^{-3\theta_{0}}$  & $e^{15\theta_{0}}$  & $e^{-15\theta_{0}}$  & $e^{3\theta_{0}}$  & 1\tabularnewline
\hline 
\end{tabular}
\par\end{center}
\end{document}